\documentclass[11pt,a4paper,epsfig]{article}

\usepackage{amsfonts}
\usepackage{graphicx}
\usepackage[colorlinks=true,linkcolor=blue]{hyperref}

\begin{document}

\title{\textbf{Kink mass quantum shifts \\from\\ SUSY quantum mechanics}}

\author{Alberto Alonso Izquierdo$^{(a)}$, Juan Mateos Guilarte$^{(b)}$, and Mikhail S. Plyushchay$^{(c)}$
\\ {\normalsize {\it $^{(a)}$ Departamento de Matematica
Aplicada and IUFFyM}, {\it Universidad de Salamanca, SPAIN}} \\ {\normalsize {\it $^{(b)}$ Departamento de Fisica
Fundamental and IUFFyM}, {\it Universidad de Salamanca, SPAIN}}\\ {\normalsize {\it $^{(c)}$ Departamento de F\'{\i}sica},
{\it Universidad de Santiago de Chile, Casilla 307, Santiago 2, CHILE}}\\ \\
\sl{\small{E-mails: alonsoiz@usal.es, guilarte@usal.es,
mikhail.plyushchay@usach.cl }}}

\date{}

\maketitle

\begin{abstract}
In this paper a new version of the DHN (Dashen-Hasslacher-Neveu) formula, which is used to compute the one-loop order
kink mass correction in (1+1)-dimensional scalar field theory models, is constructed. The new expression is written in
terms of the spectral data coming from the supersymmetric partner operator of the second-order small kink fluctuation
operator and allows us to compute the kink mass quantum shift in new models for which this calculation was
out of reach by means of the old formula.
\end{abstract}

PACS: 11.15.Kc; 11.27.+d; 11.10.Gh; 11.30.Pb

\section{Introduction}

In this work we shall bring together two research subjects in theoretical physics grown respectively in the late seventies/early eighties and developed along the past forty years. The first topic is the study of quantum fluctuations of kinks and other topological defects. Dashen, Hasslacher, and Neveu (DHN) \cite{Dashen1974} started the investigation of this problem within the framework of the $\hbar$-expansion in Quantum Field Theory  with the motivation of understanding hadrons as quantum descendants of solitary (or even solitonic) non-linear waves. Very soon after, several authors shaped this field of research almost definite both in Russia, Faddeev and Korepin \cite{Faddeev1976}, and the United States, Coleman, Jackiw, Goldstone, Christ and Lee \cite{Coleman1975,Goldstone1975,Jackiw1977,Christ1975}. The second theme is supersymmetric quantum mechanics. Introduced by Witten \cite{Witten1981} in order to elucidate a mechanism of spontaneous supersymmetry breaking, the idea triggered an inflationary process of exploring physical supersymmetric quantum mechanical systems, see the review of Cooper et al. \cite{Cooper1995} and the books of Freund \cite{Freund1986} and Junker \cite{Junker1996} to find full information about the basic ideas in SUSY quantum mechanics.

In references \cite{Dashen1975,Dashen1975b} DHN performed a very detailed analysis of the kink mass shift issue in the (1 + 1)-dimensional sine-Gordon and $\phi^4$ scalar field theory models. In particular, the authors succeeded in computing the one-loop order mass shifts for the soliton and kink non-linear waves existing in these models. They recognized that the kink mass shift induced by the quantum fluctuations must be due to three contributions: 1) the kink zero-point energy collecting the energy of the kink ground state, where all the fluctuation modes are unoccupied, 2) the analogous vacuum zero-point energy that must be subtracted, and 3) the energy induced by the one-loop renormalization mass counter-term on the kink background (measured with respect to the same effect on the vacuum). Note that the first two contributions require the computation of the difference between the traces of the kink and vacuum fluctuation operators, which are respectively Schr\"odinger and Helmholtz differential operators. The final expression encompassing all these contributions is known as the DHN (Dashen-Hasslacher-Neveu) formula and expresses the kink mass quantum shift in terms of the spectral data -bound state eigenvalues and scattering wave phase shifts- of the second-order small kink fluctuation operator. A remarkable fact to be stressed is that the kink fluctuation operators governing the $\phi^4$ and sine-Gordon kink fluctuations are reflectionless P\"oschl-Teller type Schr\"odinger operators. In these cases, both the kink and vacuum fluctuation operators exhibit a half-bound state, a singlet state with eigenvalue just at the threshold of the continuous spectrum. A general study of the (1+1) dimensional scalar field theory models enjoying this type of kink fluctuation operators can be found in the references \cite{Christ1975,Trullinger1987,Boya1989,Alonso2012}.

As early as in 1978 Olive and Witten \cite{Olive1978} unveiled the character of BPS states of the kinks (and other topological defects) in field theories with extended supersymmetry and the connection with the central charges of SUSY algebra. This property makes the BPS kink mass very robust against quantum fluctuations. Nevertheless,
by the end of the decade of the nineties the topic of quantum fluctuations of kinks underwent a comeback in the supersymmetric field theoretical framework. A thorough analysis of the DHN formula is given in \cite{Rebhan1997,Nastase1999}, where the authors uncover the mode
number cut-off regularization procedure underlying this formula. In this regularization method the difference between the traces of the kink and vacuum fluctuation operator is determined by computing the difference between the contributions of the same number of eigenmodes in the kink and vacuum fluctuation operators. The main goal of these works, however, was the quantization of supersymmetric kinks, where the contributions of fluctuations of the bosonic and the fermionic fields cancel each other, relieving the need of identifying the details of the spectrum of the kink and vacuum Hessian operators. Subtleties, however, come into play which are related to: the choice of different boundary conditions,  e.g., periodic boundary conditions, Dirichlet, and/or Robin boundary conditions, etcetera, ii) diverse regularization methods, for instance, energy cutoff, mode number cutoff, higher-order derivatives, iii) the development of phase shift analysis, see
references \cite{Shifman1999, Graham1999, Bordag2002, Goldhaber2004, Rebhan2004}.
In the $\phi^4$ and sine-Gordon models, the contributions of the half-bound states to the one-loop mass shift due to kink and vacuum fluctuation operators annihilate each other in the mode number cut-off regularization context. Thus,
the original DHN formula, which is perfectly adapted to reflectionless kink fluctuation operators, works properly. This old formula, however, needs to be slightly modified to deal with models whose
kink fluctuation operator has a non-vanishing reflection coefficient. In this case, the spectrum of kink fluctuations does not embrace a half-bound state, unlike the vacuum fluctuation operator. The necessary modification was unveiled from the one-dimensional Levinson theorem and described in reference \cite{Alonso2004}. The outcome is a generalized DHN formula, which is the starting point in this work.

The DHN formula, old or new, demands full information about the second-order small kink fluctuation
operator spectrum: not only the eigenvalues but analytical knowledge of the eigenfunctions, which encode the phase shifts, is necessary because the spectral densities enter the DHN formula. This is an important handicap in the effectiveness of these formulas since the number of exactly solvable spectral problems is reduced. Although this paper is devoted to the exact computation of the kink mass quantum correction by employing the DHN formula, we do mention the existence of alternative methods that approximately compute the one-loop kink mass shift from the spectral zeta function regularization method, see \cite{Avramidi, Elizalde1994, Kirsten2002, Vassilevich2003}. The idea is to rely on the determination of the spectral zeta function of the kink fluctuation operator via the Mellin transform of the heat trace of the same operator and then use the asymptotic expansion of this latter spectral function to obtain information about the kink fluctuation asymptotics. During the last decade two of us and our colleagues benefited from the theoretical machinery available by applying these tools at our disposal to write the one-loop kink mass shift as a truncated series in the Seeley coefficients of the
kink-Hessian-heat function in several $(1+1)$-dimensional scalar field theory models
\cite{Alonso2002,Alonso2002b,Alonso2011} and Abelian gauge Higgs models \cite{Alonso2005,Alonso2008,Mateos2008}.

In this paper, however, we shall profit from the spectral information about the kink fluctuation operator obtained by using the techniques available in supersymmetric quantum mechanics \cite{Cooper1995}. In these corresponding systems, the spectrum (eigenvalues and eigenfunctions) of a given Schr\"odinger operator is related to the spectrum of a SUSY partner operator: the two operators are iso-spectral (except possible zero modes) and the eigenfunctions are intertwined by the action of the supercharges. The spectrum of stable kink fluctuations is non-negative. Since it is the spatial derivative of the kink solution, the ground state (a zero mode)  is always known\footnote{This is due to the fact that the spatial translational symmetry of the system is spontaneously broken by the kink.}. Knowledge of the ground state wave function allows us to factorize the Schr\"odinger operator in two first-order differential operators (this being the building blocks of the supercharges) where the \lq\lq superpotential \rq\rq is the logarithm of the zero mode. The SUSY partner second-order differential operator is obtained by multiplying the two first-order operators in reverse order. By construction the SUSY partner operators are, up to zero modes, \lq\lq quasi \rq\rq-isospectral but frequently the eigenfunctions of one of the operators are simpler than the eigenfunctions of the other.

Our strategy in this work is to rewrite the DHN formula in terms of the spectral data of the new operator in the case
where the eigenfunctions of the original operators are too complicated to use the associated spectral density in the
analytical integrals. In the privileged cases of sine-Gordon, $\phi^4$, and the so-called parent models, see \cite{Alonso2012}, the SUSY partner operators describing the kink fluctuations enjoy exactly solvable spectral problems. Moreover, all these Schr\"odinger operators belong to the hierarchy of reflectionless P\"oschl-Teller potentials. Starting from a given kink Hessian one can descend through all the members of the hierarchy to end in the free Helmholtz operator. The necessary spectral data in the DHN formula applied to these kinks can be obtained by this process from the vacuum spectral data, such that  the computation of the kink mass quantum correction becomes almost trivial.

There are other models with much more complicate kink solitary waves where this SUSY quantum mechanical structure simplifies (enables) the computation of kink mass shifts due to one-loop fluctuations out of reach from a direct approach. Usually, the recursive use of the above scheme allows us to write the kink mass shift in terms of the spectral data for a SUSY partner operator that lacks bound states.  Finally, we shall construct families of field theory models whose kink fluctuation operators share a unique SUSY partner operator. Therefore, the mass quantum corrections for all the kinks in these models are identical, forming a kind of SUSY universality class.

The organization of this paper is as follows: in section 2 we introduce the notation and general background on the
subject: kink mass quantum corrections in (1+1)-dimensional scalar field theory models. In this section we also make
explicit the relations between the spectral data of the second-order small kink fluctuation operator and its
SUSY-partner operator. In subsection 2.2 we re-express the generalized DHN formula in terms of the spectral
information of this new operator, while in subsection 2.3 we accomplish the same task for any operator in
the hierarchy of SUSY-partner operators. We apply this reformulation of the DHN
formula to the sine-Gordon, $\phi^4$ and parent models, profusely studied in the literature. The novel approach
unveils the exact results in a more direct way than the standard method. Section 3 is devoted to the
application of this scheme to the computation of the kink mass quantum correction in more sophisticated scalar field
theory models. The first one is characterized by a potential energy density $U(\phi)=\frac{1}{2} \phi^2 \cos^2 \log
|\phi|$, whose topological kinks have been studied in \cite{Kumar2003}. We shall compute the kink mass quantum correction in this model by following the strategy explained above. The spectral problem of the SUSY partner operator offers much more analytical information about the eigenfunctions than the original one. In subsection 3.2 we introduce a model reconstructed from a zero mode following the procedure described in \cite{Christ1975}. In this case, the intrincate second-order small kink fluctuation operator has (miraculously) a P\"oschl-Teller type SUSY-partner operator, which again allows us to obtain the kink mass quantum shift. Finally, in Section 4 we construct families of models whose second-order small kink fluctuation operators share the same SUSY-partner operator. We shall describe in sub-Section 4.1 a family of one-gap Schr$\ddot{\rm o}$dinger operators having a unique common supersymmetric partner. In sub-Sections 4.2 and 4.3 a similar analysis will be performed on two-parametric families of two-gap
Schr$\ddot{\rm o}$dinger operators exhibiting an interesting hierarchical structure.

\section{SUSY partner operators and kink mass quantum corrections}

In this section we shall introduce the theoretical background that will be used to deduce a new version of the
Dashen-Hasslacher-Neveu (DHN) formula depending on the spectral information of the SUSY partner
to the second-order kink fluctuation operator. In some models the new formula offers us the possibility of
successfully computing the one-loop kink mass quantum correction even though it cannot be calculated directly from the standard formula.

\subsection{Field equations and topological kinks}

The action functional governing the dynamics in our (1+1)-dimensional relativistic one-scalar field theoretical models
is of the form:
\[
\tilde{S}[\psi]=\int \!\! \int\, dy^0dy^1 \, \left(\frac{1}{2}\frac{\partial\psi}{\partial y_\mu}\cdot
\frac{\partial\psi}{\partial y^\mu}- \tilde{U}[\psi(y^\mu)] \right) \quad .
\]
Here, $y^0=\tau $ and $y^1=y$ are local coordinates in ${\mathbb R}^{1,1}$, which is equipped with a metric tensor
$\eta_{\mu\nu}={\rm diag}(1,-1)$ and $\psi(y^\mu): \mathbb{R}^{1,1} \rightarrow \mathbb{R}$ is a real scalar field.

We shall work in a system of units where the speed of light is set to one, $c=1$, but we shall keep the Planck constant $\hbar$ explicit because we shall search for one-loop quantum corrections, proportional to $\hbar$, to the classical
kink masses. In this system of units, the physical dimensions are: $[\hbar]=[\tilde{S}]=M L$, $[y_\mu]=L$,
$[\psi]=M^\frac{1}{2}L^\frac{1}{2}$, $[\tilde{U}]=ML^{-1}$. The models that we shall consider are distinguished by
different choices of the part of the potential energy density that is independent of the field spatial derivatives:
$\tilde{U}[\psi(y^\mu)]$. In all of them there will be two special parameters, $m_d$ and $\gamma_d$, to be fixed
in each model, carrying the physical dimensions: $[m_d]=L^{-1}$ and $[\gamma_d]=M^{-\frac{1}{2}}L^{-\frac{1}{2}}$. We
define the non-dimensional coordinates, fields and potential in terms of these parameters: $x_\mu=m_d y_\mu$, $x_0=t$,
$x_1=x$, $\phi= \gamma_d \psi$, $U(\phi)=\frac{\gamma_d^2}{m_d^2} \tilde{U}(\psi)$. The action $\tilde{S}[\psi]$ is
also proportional to a non-dimensional action, that is, $\tilde{S}[\psi]=\frac{1}{\gamma_d^2}S[\phi]$ with
\begin{equation}
S[\phi]=\int dx^2 \left[ \frac{1}{2} \partial_\mu \phi\, \partial^\mu \phi - U(\phi) \right] \label{action} \hspace{0.4cm},
\end{equation}
where $\phi(x_\mu):\mathbb{R}^{1,1} \rightarrow \mathbb{R}$ is a non-dimensional real scalar. The second-order
non linear field equations are:
\begin{equation}
\frac{\partial^2\phi}{\partial t^2}-\frac{\partial^2\phi}{\partial x^2}=-\frac{\delta U}{\delta\phi} \label{sofe}\, \, \,  .
\end{equation}
The non-dimensional potential energy functional $\tilde{E}[\psi]=\frac{m_d}{\gamma_d^2}E[\phi]$ reads{\footnote{We shall work with non-dimensional quantities, fields, and parameters in the sequel. To recover the correct fully dimensional magnitudes we shall use the relations written above in the text.}}:
\begin{equation}
E[\phi] = \int_{-\infty}^\infty \, dx \, \varepsilon(x) = \int_{-\infty}^\infty \, dx \, \left[ \frac{1}{2} \left(
\frac{\partial \phi}{\partial x} \right)^2 + U(\phi) \right] \hspace{0.4cm} .
\label{energy}
\end{equation}
To guarantee the existence of topological kink solutions we assume that $U(\phi)$ is a non-negative twice-differentiable function, i.e., $U(\phi)\in C^2({\mathbb{R}})$ and $U(\phi)\geq 0$, with a discrete set of zeroes ${\cal M}$:
\begin{equation}
{\cal M}=\{\phi_V^{(j)} \in \mathbb{R}: U(\phi_V^{(j)})=0, \hspace{0.3cm} j=1,\dots,N\}
\label{vacs}
\hspace{0.4cm} .
\end{equation}
These time-independent and homogeneous configurations (\ref{vacs}) are the simplest solutions of the field equations
(\ref{sofe}) and the absolute minima of the energy: $E[\phi_V^{(j)}]=0$. The small  vacuum  fluctuations $\delta f_0^{(j)}(t,x)=\phi(t,x)-\phi_V^{(j)}$ are also solutions of the second-order equations (\ref{sofe}) if the linearized second-order equations are satisfied:
\begin{eqnarray}
&&\left(\frac{\partial^2}{\partial t^2}-\frac{\partial^2}{\partial x^2}+\frac{\delta^2 U}{\delta \phi^2}[\phi_V]\right)\delta f_0^{(j)}(t,x)=\left(\frac{\partial^2}{\partial t^2}-\frac{\partial^2}{\partial x^2}+v^2_j\right)\delta f_0^{(j)}(t,x)=\mathcal{O}(\delta^2) \label{lsofev} \\ && \delta f_0^{(j)}(t,x)\simeq e^{i\omega_j(k)t}f_{0k}^{(j)}(x)\, \, , \, \, K_0f_{0k}^{(j)}(x)=\left(-\frac{d^2}{dx^2}+v_j^2\right)f_{0k}^{(j)}(x)=w_j^2(k)f_{0k}^{(j)}(x) \nonumber \, \, \, .
\end{eqnarray}
The general solution of the linear PDE (\ref{lsofev}) is the Fourier integral transform
\begin{equation}
\delta f_0^{(j)}(t,x)=\frac{1}{2\pi}\int \, \frac{dk}{\sqrt{2\omega_j(k)}}\left[c_j(k)e^{i\omega_j t -i kx}+c_j^*(k)e^{-i\omega_j t +i kx}\right] \label{linwav}
\end{equation}
such that the dispersion relation $\omega_j^2(k)=k^2+v^2_j$ holds. Canonical quantization transmutes the Fourier coefficients to annihilation and creation operators, and the Fock space built from the vacuum $\phi_V^{(j)}=\langle 0_j|\hat{\phi}(t,x)|0_j\rangle$ is the space of the fundamental quanta of the system.

The configuration space ${\cal C}=\{\phi(t_0,x)\in {\rm Maps}(\mathbb{R}^{1},\mathbb{R})/E[\phi]<\infty\}$ is the union of infinite disconnected pieces: ${\rm card}({\cal M}) >1$. If this is the case, spatially extended field configuration solutions of finite energy can exist. In their center of mass they solve the second-order ODE
\begin{equation}
\frac{d^2\phi}{dx^2}=\frac{\delta U}{\delta\phi} \label{sofode}
\end{equation}
together with the asymptotic conditions
\begin{equation}
\lim_{x\rightarrow \pm \infty} \phi(x)=\phi_V^{(j_\pm)}\in {\cal M} \hspace{0.5cm},\hspace{0.5cm} \lim_{x\rightarrow \pm \infty}
\frac{\partial \phi(x)}{\partial x}=0 \hspace{0.4cm} .
\label{asymptotic}
\end{equation}
Because ${\cal M}$ is a discrete set the evolution in time from a given set of asymptotic conditions (\ref{asymptotic}) to another set would cost infinite energy. In particular, when $j_+=j_-\pm 1$ there are kink-shaped topological solutions to the ODE (\ref{sofode}), whereas if $j_-=j_+$, besides the constant solutions there might exist non-topological (bell-shaped ) kinks. The topological kinks verify the first-order ODE:
\begin{equation}
\frac{d\phi}{dx} = \pm\sqrt{2 U(\phi)} \quad .
\label{ode1}
\end{equation}
One easily proves that the solutions of this Bogomolny-Prasad-Sommerfield equation are also solutions of the ODE
(\ref{sofode}). From (\ref{ode1}) one infers that the BPS kinks are monotonically increasing functions asymptotically connecting two consecutive minima, whereas the BPS anti-kinks are monotonically decreasing functions connecting the same pair of minima.

A Lorentz transformation sends the static solution $\phi_K(x)$ to $\phi_K(t,x)=\phi_K(\frac{x-{\rm v} t}{\sqrt{1-{\rm
v}^2}})$. Thus, kinks are traveling waves that spontaneously breaks Lorentz invariance. Also, the spatial translations and reflections, $x\rightarrow x+x_0$, $x\mapsto -x$, leave the action invariant because there is no explicit dependence on $x$ in the Lagrangian, although these symmetries are spontaneously broken by the kink. Starting from one static kink $\phi_K(x)$ located at the origin, we obtain the complete family of kink/anti-kink solutions:
\begin{equation}
\phi_K(\overline{x}) \, \, , \hspace{0.5cm}  \overline{x}=(-1)^a \frac{x-x_0-{\rm v}t}{\sqrt{1-{\rm v}^2}} \, \, , \hspace{0.5cm} {\rm where} \quad
a=0,1 \quad . \label{Lcoor}
\end{equation}
Although the coordinate (\ref{Lcoor}) $\overline{x}$ should be entered  in all the formulas in the sequel regarding operators, superpotentials, kinks, etcetera, we shall always merely write $x$ (meaning that we set the kinks at their center of mass and identify kinks with anti-kinks) in order to alleviate the notation.

\subsection{Kink stability and supersymmetric quantum mechanics}

The field small fluctuations over the kink background $\delta f(t,x)=\phi(t,x)-\phi_K(x)$ are still solutions of the second-order field equations (\ref{sofe}) if the linearized PDE
\begin{equation}
\left[\Box +\frac{\delta^2 U}{\delta\phi^2}[\phi_K(x)] \right] \, \delta f(t,x)=\mathcal{O}(\delta^2)  \quad , \quad \Box=\frac{\partial^2}{\partial t^2}-\frac{\partial^2}{\partial x^2} \label{lsopdek} \quad .
\end{equation}
is satisfied. In this (static kink) case, the general solution can also be obtained
by means of separation of variables, $\delta f(t,x)=\exp[i\omega t] f_\omega(x)$, but one needs to solve the spectral problem $K_{N-}
f_\omega(x)= \omega^2 f_\omega(x)$ of the Schr\"odinger operator
\begin{equation}
K_{N-}=-\frac{d^2}{dx^2} + v_{{}_N}^2 + V_{N-}(x) \hspace{0.5cm}\mbox{where}\hspace{0.5cm} V_{N-}(x)= \frac{\delta^2
U}{\delta \phi^2}[\phi_K(x)]- v_{{}_N}^2
\label{hessianoV}
\end{equation}
instead of relying on the eigenvalues and eigenfunctions of a Helmholtz operator as in the vacuum case. Here the natural number $N\in\mathbb{N}$ refers to the number of bound states of the $V_{N-}(x)$ potential well.
We shall restrict ourselves to models  such that the function $V_{N-}(x)$ tends to zero at both $x=\pm\infty$ and, moreover, we also require:
\begin{equation}
\frac{\delta^2
U}{\delta \phi^2}[\phi_V^{(j)}]= \frac{\delta^2
U}{\delta \phi^2}[\phi_V^{(j\pm 1)}]=v_{{}_N}^2 \, \, , \, \, \quad \quad \Rightarrow \quad \quad \lim_{x\rightarrow \pm\infty} V_{N-}(x) =0 \hspace{0.4cm} .
\label{asymptoticpotential}
\end{equation}
Thus, the particle masses emerging around the two vacua connected by the kink have the same
value: $m^2=v_{{}_N}^2$. Bearing this in mind, the asymptotic behavior of (\ref{hessianoV}) is equivalent to (\ref{asymptoticpotential}){\footnote{There are interesting models where the vacua connected by the kink are not equivalent, giving rise to particles with different masses, e.g., the Khare-Lohe $\phi^6$ model. One understands that the moduli space of vacua is formed by several orbits of the symmetry group. The quantum mechanical spectral problem posed by the kink Hessian in this situation is non unitary.}}.

The stability of the solution $\phi_K$ is guaranteed when the eigenvalues of the operator (\ref{hessianoV}) are
non-negative, because in this case the fluctuations $\delta f(t,x)$ remain bounded in the temporal evolution. In general, the spectrum of the operator (\ref{hessianoV}) is of the form
\begin{equation}
{\rm Spec}(K_{N-})=\{ \omega_n^2\}_{n=0,1,\dots,l-1} \cup \{\omega_l^2\}_{s_l} \cup \{k^2+v_{{}_N}^2\}_{k\in \mathbb{R}}
\hspace{0.4cm} , \hspace{0.4cm} \omega_n^2<\omega_{n+1}^2 \, \, ,
\label{spectrum}
\end{equation}
where $\omega_n^2$ are the bound state eigenvalues, $l+1$ is the number of eigenfunctions in the discrete spectrum, and $\omega^2(k)=k^2+v_{{}_N}^2$
correspond to the continuous spectrum eigenvalues of the $K_{N-}$ operator. According to the one-dimensional Levinson theorem, the subscript $s_l$ in (\ref{spectrum})
to be entered in the DHN formula, distinguishes whether the highest eigenvalue in the discrete spectrum, $\omega_l^2$, comes from a bound or a half-bound state. If
$\omega_l^2 < v_{{}_N}^2$ this eigenvalue corresponds to a (normalizable) bound state, $N=l+1$, and we shall choose $s_l=1$ in the DHN formula, whereas if
$\omega_l^2=v_{{}_N}^2$ there is a (non-normalizable but bounded at infinity) half-bound state just at the threshold of the continuous spectrum; thus $N=l$, and the DHN choice will be $s_l=\frac{1}{2}$.

To show the kink stability we differentiate with respect to $x$ the ODE equation (\ref{sofode}) evaluated at the kink solution:
\begin{equation}
\frac{d}{dx}\cdot\frac{d^2\phi_K}{dx^2}=\frac{d}{dx}\cdot \frac{\delta U}{\delta\phi}[\phi_K(x)]\, \, \equiv \, \,
-\left(-\frac{d^2}{dx^2}+\frac{\delta^2 U}{\delta\phi^2}[\phi_K(x)]\right)\frac{d\phi_K}{dx}=0 \label{sofodezm} \quad .
\end{equation}
The wave function
\begin{equation}
\psi_0(x)=\frac{d\phi_K}{dx}(x)
\label{zeromode}
\end{equation}
is accordingly an eigenfunction of the $K_{N-}$ operator with eigenvalue zero. For topological kinks, monotonic functions of $x$, their derivatives have no nodes and the translational mode $\frac{d\phi_K}{dx}$ is the ground state of the second-order kink fluctuation operator, i.e., $\omega_0^2=0$, and topological kinks are always stable. Non-topological kinks, however, give rise to translational modes with one node and cannot be the ground state. There is an eigenstate of negative energy and non-topological kinks are unstable. Thus, the one-loop mass shift of non-topological kinks picks an imaginary part since it corresponds to resonances in the quantum domain, whereas stable topological kinks are bona fide quantum eigenstates of the field theoretical Hamiltonian.

The existence of the zero mode ground state is the crucial fact for building the structure of supersymmetric quantum mechanics in the $K_{N-}$ spectral problem. The $K_{N-}$ operator can be written as the product of two first-order operators in the form
\[
K_{N-}=A_N^\dagger A_N\hspace{0.5cm}\mbox{with} \hspace{0.5cm}
A_N=\frac{d}{dx}+W_N(x) \hspace{0.3cm} \mbox{and} \hspace{0.3cm} A_N^\dagger = -\frac{d}{dx} +W_N(x) \hspace{0.4cm} ,
\]
where the function $W_N(x)$ is minus the derivative of the superpotential $\Xi_N(x)=\log\psi_0(x)$, the logarithm of
the ground state wave function:
\begin{equation}
W_N(x)=-\Xi_N^\prime (x)=-\frac{\psi_0^\prime(x)}{\psi_0(x)} \hspace{0.4cm} .
\label{superpotential}
\end{equation}
Explicitly, the second-order kink fluctuation $K_{N-}$ operator is defined in terms of the $W_N(x)$ function:
\begin{equation}
K_{N-}=-\frac{d^2}{dx^2} + W_N(x)^2-W_N^\prime(x) \hspace{0.4cm} , \hspace{0.4cm}  W_N(x)^2-W_N^\prime(x)=v_{{}_N}^2+V_{N-}(x) \quad .
\label{Hessian00}
\end{equation}
Quite generically the asymptotic behavior of the derivative of the superpotential at very large distances from the origin is :  $W_N(x)
\approx_{x\to \pm \infty} \, \, (-1)^{\alpha} v_{{}_N} \tanh[v_{{}_N} x]$ where $\alpha=0,1$. Thus,
\begin{equation}
\lim_{x\rightarrow \pm \infty} W_N(x) =\pm (-1)^\alpha v_{{}_N} \hspace{0.5cm} , \hspace{0.5cm} \lim_{x\to \pm \infty}
W_N^\prime(x)  =0 \hspace{0.4cm} .
\label{asymptoticsuperpotential}
\end{equation}
There is a partner Schr\"odinger operator that is also factorized in terms of the first-order $A_N$ and $A_N^\dagger$ operators but in the reverse order:
\begin{equation}
K_{N+}=A_NA_N^\dagger =-\frac{d^2}{dx^2} + W_N(x)^2 +W_N^\prime(x)=-\frac{d^2}{dx^2} + v_{_{N}}^2 +V_{N+}(x)\, \, \, .
\label{partneroperator}
\end{equation}

All this calls for the construction of the following $\mathcal{N}=2$ supersymmetric quantum mechanical system: there are two supercharges
\begin{equation}
\hat{Q}_N=\left(\begin{array}{cc} 0 & 0 \\ A_N & 0\end{array}\right) \hspace{1cm} , \hspace{1cm} \hat{Q}_N^\dagger=\left(\begin{array}{cc} 0 & A_N^\dagger \\ 0 & 0\end{array}\right)
\label{supchar}
\end{equation}
that close the supersymmetry algebra
\begin{equation}
\hat{Q}_N^2=(\hat{Q}_N^\dagger)^2=0 \hspace{0.5cm} , \hspace{0.5cm}  \hat{K}_N=\hat{Q}_N\hat{Q}_N^\dagger+\hat{Q}_N^\dagger\hat{Q}_N \label{susalg} \quad .
\end{equation}
The Hamiltonian
\[
\hat{K}_N=\left(\begin{array}{cc} K_{N-} & 0 \\ 0 & K_{N+}\end{array}\right) =\left(\begin{array}{cc}  -\frac{d^2}{dx^2} + W_N(x)^2 -W_N^\prime(x) & 0 \\ 0 & -\frac{d^2}{dx^2} + W_N(x)^2 +W_N^\prime(x)\end{array}\right)
\]
is supersymmetric because it commutes with the supercharges, $\hat{K}_N\hat{Q}_N-\hat{Q}_N\hat{K}_N=\hat{K}_N\hat{Q}_N^\dagger-\hat{Q}_N^\dagger\hat{K}_N=0$, which are consequently constants of motion of the system. In sub-Section 2.4 we shall describe  the very well known properties of the spectrum of supersymmetric Hamiltonians that will help us in our main task of computing one-loop kink mass shifts.

\subsection{SUSY quantum mechanics, the BPS equation, and reconstruction of scalar field models supporting kinks}
We now shall explain how the supersymmetric quantum mechanical structure of the topological kink stability problem is indeed deeply encrypted in the BPS or first-order ODE (\ref{ode1}){\footnote{This structure has been derived in the previous sub-Section from the existence of the translational zero mode.}}. Plugging the field configurations written as the kink plus a small fluctuation $\phi(x)=\phi_K(x)+\delta\phi(x)$ into the BPS equations we find new solutions if and only if the following linear ODE holds:
\begin{equation}
\left(\frac{d}{dx}\mp\frac{U^\prime(\phi_K)}{\sqrt{2U(\phi_K)}}\right)\delta\phi(x)=0 \label{folde} \qquad .
\end{equation}
Because the topological kink solves equation (\ref{sofode}) we know that $U^\prime(\phi_K)=\psi_0^\prime$ and the equation (\ref{ode1}) means that $\sqrt{2 U(\phi_K)}=\psi_0$. The
differential operators in the equation (\ref{folde}) are no more than the factorization operators $A_N$ and $A_N^\dagger$. Therefore, the factorization property of the second-order
kink fluctuation operator is closely related to the existence of first-order equations in the scalar field theory model.

This, at first sight, hidden property of BPS kinks is in the backyard of the mechanism essentially discussed in reference \cite{Trullinger1987} and used explicitly in papers \cite{Kumar1987,Dey1994}. Starting from a given superpotential function $\Xi_N(x)$ the field theory model can be reconstructed
by following the subsequent steps{\footnote{Another interesting procedure of reconstruction based in the inverse scattering method has been proposed on \cite{Bordag2003}.}}: (a) From the equation (\ref{superpotential}) we write the zero mode in terms of the superpotential:
\begin{equation}
\psi_0(x)=e^{\Xi_N(x)}
=e^{-\int^x d\xi W_N(\xi)} \qquad .
\label{zeromodefromsuperpotential}
\end{equation}
(b) From the zero mode (\ref{zeromode}) the kink solution is obtained through indefinite spatial integration:
\begin{equation}
\phi_K(x)= \int^x \psi_0(\xi) d\xi \qquad .
\label{kinkfromzeromode}
\end{equation}
(c)  The final step is the most delicate one. The first-order equation (\ref{ode1}) allows us to write the function $U(\phi) $ characterizing the field theory model
\begin{equation}
U(\phi)=\frac{1}{2} \psi_0^2(\phi_K^{-1}(\phi)) \hspace{0.4cm} ,
\label{potentialfromzeromode}
\end{equation}
from the inverse of the bijective function $\phi=\phi_K(x)$, e.g., $x=\log{\rm tan}\, \frac{\phi}{4}$ for the sG-kink, or $x={\rm arctanh\, \phi}$ for the $\phi^4$ kink. Formula (\ref{potentialfromzeromode}) defines $U(\phi)$ in the range of the kink $\phi_K(x)$, i.e., in the interval $\phi(t,x)\in [\phi_V^{(j)},\phi_V^{(j+1)}]$. Some type of ${\cal C}^2(\mathbb{R})$-continuity is needed to extend $U(\phi)$ to the rest of the field space.  We shall make use of this procedure in the section 3 in order to construct some field theoretical models starting from a given kink zero mode.
\subsection{Kink mass quantum correction from the SUSY partner operator}
After identifying a classical kink solution we are interested in analyzing its quantum properties. Pursuing this goal
Dashen, Hasslacher and Neveu succeeded in computing the one-loop quantum correction to the classical mass of the kink
and soliton found in the $\phi^4$ and sine-Gordon models \cite{Dashen1974,Dashen1975,Dashen1975b}. After a zero-point
renormalization and a subsequent mass renormalization, these authors obtained what is now known as the DHN
formula, which has been generalized in \cite{Alonso2004} for operators having a non-transparent scattering spectrum. The generalized DHN formula applicable both to transparent and non-transparent Hamiltonians reads:
\begin{equation}
\frac{\Delta E(\phi_K)}{\hbar\gamma_d^2}= \frac{1}{2}\sum_{n=0}^{l-1} \omega_n + \frac{1}{2} s_{l} \omega_{l}-
\frac{v_{{}_N}}{4}  + \frac{1}{4\pi} \left<V_{N-}(x) \right> + \frac{1}{2\pi} \int_0^\infty dk \left[ \frac{\partial
\delta^{(N-)}(k)}{\partial k} \sqrt{k^2 + v_{{}_N}^2} - \frac{\frac{1}{2}  \left<V_{N-}(x) \right>}{\sqrt{k^2+v_{{}_N}^2}}
\right]
\label{gdhn}
\end{equation}
where $\omega_n^2$ are the bound state eigenvalues of the second-order kink fluctuation $K_{N-}$ operator. $s_l$ is
equal to $\frac{1}{2}$ if the spectrum of the $K_{N-}$ operator involves one half-bound eigenfunction and it is one
otherwise{\footnote{We recall that a half-bound state is a singlet eigenstate of energy eigenvalue equal to the energy of the threshold continuous spectrum. The reason for the $\frac{1}{2}$ factor (and the name) is the one-dimensional Levinson theorem.}}. Moreover, $\delta^{(N-)}(k)$ denotes the phase shifts in the scattering waves of the $K_{N-}$ operator and $\left< V_{N-}(x) \right>$ stands for the area enclosed by the function $V_{N-}(x)$ and the real line: $\left< V_{N-}(x)
\right>=\int_{-\infty}^\infty V_{N-}(x) dx$. We recall that $V_{N-}(x)$ is the potential well of $K_{N-}$ minus the
non-zero threshold $v_{{}_N}^2$. It is worthwhile at this point to explain the physical origin of the different terms in formula (\ref{gdhn}):
\begin{enumerate}

\item The first two terms encode the contribution of the singlet (bound and half-bound) states to the kink fluctuation energy. The third term corresponds to the subtraction of the singlet state of the vacuum fluctuation spectrum.

\item The positive term in the integral over the scattering modes  collects all the kink minus vacuum continuous spectrum fluctuations to the energy.

\item The finite fourth term comes from using an intermediate regularization setting the same number of kinks as
vacuum fluctuation modes, rather than a cutoff in the kink and vacuum energies. All these terms assemble together make what one might call kink Casimir energy.

\item The remaining, negative, term in the integral is due to the contribution to the fluctuation energy of the
one-loop renormalization mass counterterm.

\end{enumerate}

Nevertheless, the use of expression (\ref{gdhn}) brings with it a handicap: complete analytical knowledge of the spectral data concerning the second-order kink fluctuation $K_{N-}$ operator is a rare occurrence in generic (1+1)-dimensional scalar field models. This makes the computation of the kink mass shift
seldom feasible. The goal of this paper is to take advantage of the techniques available in 1D supersymmetric quantum mechanics in order to alleviate the dependence on the spectral information of the DHN formula (\ref{gdhn}).

The aim in this section is to rewrite the generalized DHN formula (\ref{gdhn}) in terms of the spectral data
of the supersymmetric partner operator $K_{N+}$ to $K_{N-}$. The supersymmetric algebra shows that
the spectral data concerning the $K_{N-}$ operator, defined in (\ref{hessianoV}) and
(\ref{Hessian00}), and the data of its supersymmetric partner $K_{N+}$, defined in (\ref{partneroperator}),
are closely related \cite{Cooper1995}. First, bearing in mind that $V_{N+}(x)=W_N(x)^2+W_N^\prime(x)-v_N^2$ we find that
\[
\lim_{x\rightarrow \pm \infty} V_{N+}(x)=0 \hspace{0.4cm} ,
\]
showing that the $K_{N-}$ and $K_{N+}$ operators share the threshold $v_N^2$ of the continuous spectrum, see
Figure 1. Second, if the spectrum of the $K_{N-}$ operator is summarized in the form (\ref{spectrum}) then
the spectrum of $K_{N+}$ reads:
\[
{\rm Spec }\, K_{1+} = \{\omega_n^2\}_{n=1,\dots,l-1} \cup \{\omega_l^2\}_{s_l} \cup \{k^2+v_N^2\}_{k\in \mathbb{R}}
\hspace{0.4cm} .
\]
Both spectra look identical but there are two important differences: (1) In ${\rm Spec }\, K_{1+}$ there is no zero
mode. (2) Although the continuous spectrum eigenvalues are the same the spectral densities are not.
\begin{figure}[ht]
\centerline{\includegraphics[height=4cm]{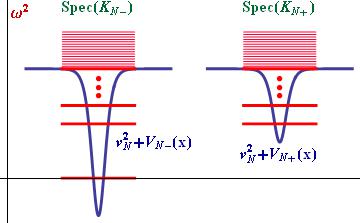} }

\caption{\small Graphical representation of the spectra of the kink fluctuation operator $K_{N-}$ and its SUSY partner
operator $K_{N+}$ (red horizontal lines) and the potential well arising in these operators (solid blue lines).}
\end{figure}

Third, the phase shifts emerging in the scattering eigenfunctions of these two $K_{N+}$ and $K_{N-}$ operators are also related \cite{Cooper1995}.
Given the asymptotic behavior of the right-going scattering waves of $K_{N-}$ and $K_{N+}$
\begin{eqnarray*}
&&\psi^{(N-)}(k,x\rightarrow -\infty)\rightarrow e^{ik x} + \rho_{N-}^{(\rightarrow)}e^{-ik x}
\hspace{0.5cm},\hspace{0.5cm} \psi^{(N-)}(k,x\rightarrow \infty) \rightarrow\sigma_{N-}^{(\rightarrow)} e^{ik x} \hspace{0.4cm}
\\
&&\psi^{(N+)}(k,x\rightarrow -\infty)\rightarrow e^{ik x} + \rho_{N+}^{(\rightarrow)} e^{-ik
x}\hspace{0.5cm},\hspace{0.5cm} \psi^{(N+)}(k,x\rightarrow \infty) \rightarrow  \sigma_{N+}^{(\rightarrow)} e^{ik x}
\hspace{0.4cm} ,
\end{eqnarray*}
where $k^2= \omega^2-v_N^2$, the relations between the $K_{N\pm}$ reflection and transmission scattering amplitudes are:
\[
\rho_{N-}^{(\rightarrow)} = \frac{W_{N-}+ik}{W_{N-} -ik}\,\rho_{N+}^{(\rightarrow)} \hspace{0.3cm},\hspace{0.3cm}
\sigma_{N-}^{(\rightarrow)}=\frac{W_{N+}-ik}{W_{N-} -ik}\, \sigma_{N+}^{(\rightarrow)} \hspace{0.3cm},\hspace{0.3cm}
\mbox{ where } W_{N\pm} = \lim_{x\rightarrow \pm \infty} W_N(x) \hspace{0.4cm} .
\]
Analogously, for an incident wave coming from the right side we have the relations:
\[
\rho_{N-}^{(\leftarrow)}=\frac{W_{N+}-ik}{W_{N+}+ik}\, \rho_{N+}^{(\leftarrow)} \hspace{0.3cm},\hspace{0.3cm}
\sigma_{N-}^{(\leftarrow)}=\frac{W_{N-}+ik}{W_{N+}+ik}\, \sigma_{N+}^{(\leftarrow)} \hspace{0.4cm} .
\]
Without loss of generality we choose $W_{N+}=v_{{}_N}=-W_{N-}$. In this situation $\sigma_{N+}^{(\leftarrow)}
=\sigma_{N+}^{(\rightarrow)} =\sigma_{N+}$ and $\sigma_{N-}^{(\leftarrow)}=\sigma_{N-}^{(\rightarrow)}=\sigma_{N-}$.
The identities immediately above reduce to
\begin{equation}
\rho_{N-}^{(\rightarrow)} = \frac{v_{{}_N}-ik}{v_{{}_N}+ik}\rho_{N+}^{(\rightarrow)} \hspace{0.3cm},\hspace{0.3cm}
\rho_{N-}^{(\leftarrow)}=\frac{v_{{}_N}-ik}{v_{{}_N}+ik}\rho_{N+}^{(\leftarrow)} \hspace{0.3cm},\hspace{0.3cm}
\sigma_{N-}=-\frac{v_{{}_N}-ik}{v_{{}_N}+ik}\sigma_{N+} \hspace{0.4cm} .
\label{coefficients}
\end{equation}
The unitary scattering matrix $S_{N-}$ encoding the scattering data of the operator $K_{N-}$ can also be defined in terms of the reflection and transmission scattering amplitudes of the supersymmetric partner operator
\[
S_{N-}=\left( \begin{array}{cc} \sigma_{N-} & \rho_{N-}^{(\leftarrow)} \\ \rho_{N-}^{(\rightarrow)} & \sigma_{N-}
\end{array} \right) = \frac{v_{{}_N} - ik}{v_{{}_N} + ik} \left( \begin{array}{cc}  - \sigma_{N+} & \rho_{N+}^{(\leftarrow)}
\\[0.3cm]   \rho_{N+}^{(\rightarrow)}  &  -\sigma_{N+} \end{array} \right) \hspace{0.4cm} .
\]
The phase shifts $\delta_\pm^{(N-)}$ are the arguments of the eigenvalues
$\lambda_\pm^{(N-)} = e^{2i\delta_{\pm}^{(N-)}}$ of the matrix $S_{N-}$.  Thus, the relation between the phase shifts of the supersymmetric $K_{N-}$ / $K_{N+}$ operator pair follows:
\begin{equation}
e^{2i\delta_\pm^{(N-)}} = \sigma_{N-} \pm \sqrt{\rho_{N-}^{(\leftarrow)} \rho_{N-}^{(\rightarrow)}} =  - \frac{v_{{}_N} -
ik}{v_{{}_N} + ik} \left[\,\sigma_{N+} \mp \sqrt{\rho_{N+}^{(\leftarrow)} \rho_{N+}^{(\rightarrow)}} \,\right] = -\frac{v_{{}_N}
- ik}{v_{{}_N} + ik} e^{2i\delta_\mp^{({N+})}} \quad .
\label{phase1}
\end{equation}
Defining a relative phase shift
$\theta_0$ by the formula $e^{i\theta_0} = \frac{v_{{}_N} -ik}{v_{{}_N} + ik}$, i.e.,
\begin{equation}
\theta_0(k)=\arctan \frac{2 v_{{}_N} k}{k^2-v_{{}_N}^2} \hspace{0.4cm} ,
\label{relativephase}
\end{equation}
expression (\ref{phase1}) becomes
\[
\delta_\pm^{(N-)}(k) = \delta_\mp^{({N+})}(k) + \frac{1}{2}\theta_0(k) + \frac{\pi}{2} \hspace{0.4cm} .
\]
Therefore, the connection between the $K_{N\pm}$ total phase shifts $\delta^{(N\pm)}(k) =
\delta_+^{(N\pm)}(k)+\delta_-^{(N\pm)}(k)$  is
\[
\delta^{(N-)}(k) = \delta^{({N+})}(k) + \theta_0(k) + \pi \hspace{0.4cm} .
\]
The generalized DHN formula (\ref{gdhn})  requires the derivative of this phase shifts with respect to the momentum $k$, which is the spectral density of the kink fluctuation minus the spectral density of the vacuum fluctuation,
\[
\frac{d\delta^{(N-)}(k)}{dk} = \frac{d\delta^{({N+})}(k)}{dk} -\frac{2v_N}{k^2+v_{{}_N}^2} \hspace{0.4cm} ,
\]
where we have used (\ref{relativephase}). Plugging this result into (\ref{gdhn}) we obtain a new version of the
generalized DNH formula
\begin{equation}
\frac{\Delta E(\phi_K)}{\hbar\gamma_d^2}= \frac{1}{2}\sum_{n=1}^{l-1} \omega_n + \frac{1}{2} s_{l} \omega_{l}-
\frac{v_{{}_N}}{4}  + \frac{1}{4\pi} \left<V_{N-}(x) \right> + \frac{1}{2\pi} \int_0^\infty dq \left[  \frac{d
\delta^{({N+})}(k)}{dk}\sqrt{k^2 + v_{{}_N}^2} - \frac{\frac{1}{2}  \left<V_{N-}(x) \right>+ 2v_{{}_N}}{\sqrt{k^2 +
v_{{}_N}^2}}\right]
\label{gdhn2}
\end{equation}
Fortunately the $\delta^{(N\pm)}(k)$ phase shifts are related by supersymmetry in such a way that the modification of the DHN formula merely requires a shift in the mass renormalization contribution to the kink fluctuation energy.
The new formula depends on the spectral data of the supersymmetric partner operator $K_{N+}$ instead of $K_{N-}$
itself. Sometimes the analysis of the spectrum of $K_{N+}$ is simpler than that of $K_{N-}$ and the $\delta^{({N+})}(k)$ phase shifts are analytically accessible, whereas the $\delta^{({N-})}(k)$ either
are not accessible or are much more involved. In this situation formula (\ref{gdhn2}) is more manageable than the original generalized DHN formula (\ref{gdhn}).

\subsubsection{One-loop mass correction from the sine-Gordon soliton fluctuations}

We shall illustrate the use of (\ref{gdhn2}) by means of its application to the computation of the one-loop quantum
correction of the sine-Gordon soliton mass. In this model we need only the $N=1$ version of the general formalism. Use of formula (\ref{gdhn2}) renders this calculation almost trivial. The sine-Gordon potential is
\begin{equation}
U(\phi)=1-\cos \phi \hspace{0.4cm} .
\label{SineGordonPotential}
\end{equation}
The degenerate absolute minima of $U(\phi)$ form the discrete set ${\cal M}=\{\phi^{(j)}=2\pi j\}$, $j\in\mathbb{Z}$, and hence the masses of the fundamental quanta are: $v_1^2=\frac{\delta^2U}{\delta\phi^2}\left.\right|_{\phi^{(j)}}=1, \forall j$.
Besides these static homogeneous solutions the model admits the kink extended solutions
\[
\phi_S(x)=\pm 4 \arctan e^{x}+2\pi j \quad ,
\]
which interpolate respectively between the $\phi^{(j)}=2\pi j$ and  $\phi^{(j\pm 1)}= 2\pi (j \pm 1)
$ minima. The second-order sG-soliton fluctuation operator (\ref{hessianoV}) is
\begin{equation}
K_{1-}=-\frac{d^2}{dx^2}+1-2\,{\rm sech}^2\,x \hspace{0.4cm} ,
\label{solitonhessian}
\end{equation}
which fixes $V_{1-}(x)=-2\,{\rm sech}^2\,x$ and therefore $\left<V_{1-}(x)\right>=-4$. Following the scheme described
in the section 2.1 we find from (\ref{zeromode}) that the zero mode of $K_{1-}$ is: $\psi_0(x)=2\,{\rm sech}\, x$. Also this Hamiltonian admits a half-bound state: $\psi_{1/2}(x)=\tanh x$ of energy one.
Therefore, the superpotential and its derivative are: $\Xi_1(x)=-\log\cosh x$, $W_1(x)=\tanh x$, as one may see from (\ref{superpotential}). The
$A_1=\frac{d}{dx}+\tanh x$ first-order operator permits the factorization of (\ref{solitonhessian}) in the form $K_{1-}=A_1^\dagger A_1$ and unveils its
supersymmetric partner (\ref{partneroperator}) :
\[
K_{1+}=A_1A_1^\dagger =-\frac{d^2}{dx^2}+1 \hspace{0.4cm} .
\]
Obviously $V_{1+}(x)=0$ and the phase shift is null: $\delta^{(1+)}(k)=0$. The
spectrum of the $K_{1+}$ operator includes the continuous spectrum of plane waves and energy $\omega^2(k)=k^2+1$ plus one singlet (half-bound) state, a constant function, of
energy one. Therefore, there is only one bound state in the spectrum of the SUSY pair: the $K_{1-}$  zero mode. The one-loop sG-soliton mass quantum correction is computed by applying the formula (\ref{gdhn2}) to find
\[
\frac{\Delta E(\phi_K)}{\hbar \gamma_d^2} = \frac{1}{4\pi}(-4) + \frac{1}{2\pi} \int_0^\infty dk \left[ 0 -
\frac{-2+2}{\sqrt{k^2+1}} \right] = -\frac{1}{\pi} \hspace{0.4cm} ,
\]
which recovers the classical result of Dashen, Hasslacher and Neveu \cite{Dashen1974,Dashen1975} in an elementary way.

\subsection{Kink mass quantum corrections from SUSY quantum mechanical hierarchies }

In the previous section we profited from the existence
of a supersymmetric partner $K_{N+}$ to the second-order kink fluctuation operator $K_{N-}$ in order to obtain a new DHN formula (\ref{gdhn2}), which depends only on the
spectral data of $K_{N+}$. We shall continue exploiting this strategy by building a SUSY quantum mechanical hierarchy that will allow us to reduce the kink mass computation in terms of the spectral data of simpler Hamiltonians in many more models. Starting from the (as yet unspecified but factorized)  operator $K_{N+}=A_N A_N^\dagger$ we define  a new operator as
\[
K_{(N-1)-} = K_{N+}-u_N^2 = -\frac{d^2}{dx^2} + v_N^2-u_N^2 +V_{N+}(x) =  -\frac{d^2}{dx^2} + v_{(N-1)}^2 +V_{(N-1)-}(x)
\hspace{0.4cm} ,
\]
where $v_{(N-1)}^2=v_N^2-u_N^2$ and $V_{(N-1)-}(x)=V_{N+}(x)$, as the starting operator of a new stage. If the $K_{(N-1)-}$ operator admits the factorization
\[
K_{(N-1)-}=A_{(N-1)}^\dagger A_{(N-1)}  \hspace{0.5cm}\mbox{with} \hspace{0.5cm}
A_{(N-1)}=\frac{d}{dx}+W_{(N-1)}(x) \hspace{0.3cm} \mbox{and} \hspace{0.3cm} A_{(N-1)}^\dagger = -\frac{d}{dx} +W_{(N-1)}(x) \hspace{0.4cm} ,
\]
the strategy explained in the previous section can be applied iteratively. This happens with the choice $u_N^2=\omega_1^2$, i.e., we subtract the energy of the first excited level of $K_{N-}$ identical to the ground state
energy of $K_{N+}$. In terms of the new function
$W_{(N-1)}(x)$ the $K_{(N-1)-}$ operator reads:
\[
K_{{(N-1)}-}=-\frac{d^2}{dx^2} + W_{(N-1)}(x)^2 - W_{(N-1)}^\prime(x) \hspace{0.4cm} .
\]
Taking into account that the definition of $K_{(N-1)-}$ only involves a translation
of the $K_{N+}$ origin of energies, the spectra of $K_{(N-1)-}$ and $K_{N+}$ are related trivially. Denoting
\[
{\rm Spec}\,K_{(N-1)-} = \{ \nu_n^2\}_{n=1,\dots,l-1} \cup \{\nu_l^2\}_{s_l} \cup \{k^2+v_{(N-1)}^2\}_{k\in \mathbb{R}}
\]
the relation $\nu_n^2=\omega_n^2- \omega_1^2=\omega_n^2-(v_N^2-v_{(N-1)}^2)$ shows that $\nu_1^2=0$ and the kernel of $K_{(N-1)-}$ exhibits a zero mode. The phase shifts associated with the $K_{(N-1)-}$ and $K_{N+}$ operators obviously coincide: $\delta^{(N-1)-}(k)=\delta^{N+}(k)$.

\begin{figure}[ht]
\centerline{\includegraphics[height=4cm]{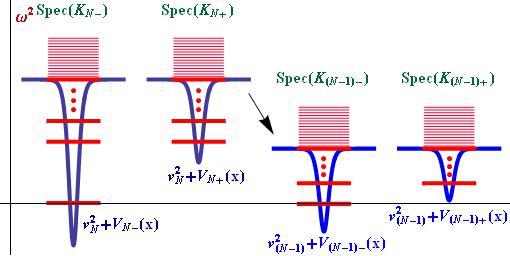} }

\caption{\small Graphical representation of the energy spectra of the SUSY-partner operator hierarchy $K_{N\mp}$, $K_{(N-1)\mp}$, (red horizontal lines) and the potential wells characterizing these operators (solid blue lines).}
\end{figure}

The following step is to construct the supersymmetric partner operator to $K_{(N-1)-}$
\[
K_{(N-1)+} = A_{(N-1)} A_{(N-1)}^\dagger = -\frac{d^2}{dx^2} + W_{(N-1)}(x)^2 + W_{(N-1)}^\prime(x) = -\frac{d^2}{dx^2} + v_{(N-1)}^2 +V_{(N-1)+}(x) \hspace{0.4cm}
,
\]
where the notation $V_{(N-1)+}(x)=W_{(N-1)}(x)^2 + W_{(N-1)}^\prime(x) - v_{(N-1)}^2$ is evident see Figure 2. Again, there exists a connection between the spectral data of the $K_{(N-1)-}$ and $K_{N+}$ operators:
\[
{\rm Spec}\,K_{(N-1)+} = \{ \nu_n^2\}_{n=2,\dots,l-1} \cup \{\nu_l^2\}_{s_l} \cup \{k^2+v_{(N-1)}^2\} \, \, \, ,
\]
note the absence of the zero mode, and the $\delta^{((N-1)+)}(k)$ phase shift derivatives satisfy:
\begin{eqnarray*}
\frac{d\delta^{((N-1)+)}(k)}{dk} &=& \frac{d\delta^{((N-1)-)}(k)}{dk} + \frac{2v_{(N-1)}}{k^2+v_{(N-1)}^2}=  \frac{d\delta^{(N+)}(k)}{dk} +
\frac{2v_{(N-1)}}{k^2+v_{(N-1)}^2}=\\ &=& \frac{d\delta^{(N-)}(k)}{dk} + \frac{2v_N}{k^2+v_N^2} + \frac{2v_{(N-1)}}{k^2+v_{(N-1)}^2} \hspace{0.4cm}.
\end{eqnarray*}
Plugging this last result into formula (\ref{gdhn2}) the generalized DHN formula is promoted to this level of the
hierarchy:
\begin{eqnarray}
\frac{\Delta E(\phi_K)}{\hbar\gamma_d^2} &=& \frac{1}{2}\sum_{n=1}^{l-1} \omega_n + \frac{1}{2} s_{l} \omega_{l}-
\frac{v_N}{4}  + \frac{1}{4\pi} \left<V_{N-}(x) \right> + \nonumber  \\ && +\frac{1}{2\pi} \int_0^\infty dk \left[
\frac{d \delta^{((N-1)+)}(k)}{dk}\sqrt{k^2 + v_N^2} - \frac{\frac{1}{2}  \left<V_{N-}(x) \right>+ 2v_N}{\sqrt{k^2 + v_{(N-1)}^2}}
- \frac{2v_{(N-1)}\sqrt{k^2 + v_N^2}}{k^2 + v_{(N-1)}^2} \right] \hspace{0.4cm} . \label{gdhn3}
\end{eqnarray}
Note that this depends on the spectral information of the $K_{(N-1)+}$ operator.

Formula (\ref{gdhn3}) can be further generalized by repeating these steps in a recursive way. By induction it is easy to check that the phase shift derivative arising in the $K_{(N-j)+}$ operator, the $j$-th level of the supersymmetric hierarchy, satisfies the condition
\[
\frac{d\delta^{((N-j)+)}(k)}{dk} = \frac{d\delta^{((N-j)-)}(k)}{dk} + \frac{2v_{(N-j)}}{k^2+v_{(N-j)}^2} = \frac{d\delta^{(N-)}(k)}{dk}
+ \sum_{i=0}^{j} \frac{2v_{(N-i)}}{k^2+v_{(N-i)}^2} \hspace{0.4cm} ,
\]
where $v_{(N-i)}^2=v_N^2-\omega_i^2$. In this level of the hierarchy the generalized DHN formula
\begin{eqnarray}
\frac{\Delta E(\phi_K)}{\hbar\gamma_d^2} &=&
\frac{1}{2}\sum_{n=1}^{l-1} \omega_n + \frac{1}{2} s_{l} \omega_{l}-
\frac{v_N}{4}  + \frac{1}{4\pi} \left<V_{N-}(x) \right> + \nonumber
\\ && +\frac{1}{2\pi} \int_0^\infty dk \left[ \frac{d
\delta^{((N-j)+)}(k)}{dk}\sqrt{k^2 + v_N^2} - \frac{\frac{1}{2}
\left<V_{N-}(x) \right>+ 2v_N}{\sqrt{k^2 + v_N^2}} - \sum_{i=1}^{j}
\frac{2v_{(N-i)}\sqrt{k^2 + v_N^2}}{k^2 + v_{(N-i)}^2} \right]
\hspace{0.4cm} \,\, \label{gdhn4}
\end{eqnarray}
expresses the kink mass quantum shift in terms of the spectral data of the $j$-th supersymmetric
operator with trivial kernel.

\subsubsection{One-loop quantum correction to the $\phi^4$-kink mass}

The $\phi^4$ model is characterized by the potential energy density $U(\phi) =\frac{1}{2} (\phi^2-1)^2$ in
the action functional (\ref{action}). The set of vacua is ${\cal M}=\{-1,1\}$. The fundamental quanta emerging from any of these two vacua have masses equal to $v_2^2=4$. The kink solutions
\[
\phi_K(x)= \tanh x
\]
connect the two vacua of the model. In this case $N=2$
because the $\phi^4$ kink second-order fluctuation operator presents two bound states:
\[
K_{2-}=\frac{d^2}{dx^2}+4-6\,{\rm sech}^2\, x \hspace{0.4cm} .
\]
The potential well $V_{2-}(x)=-2(2+1)\,{\rm sech}^2\,x$ is the second in the hierarchy of
 P\"oschl-Teller reflectionless potentials. Moreover, $\left< V_{2-}(x) \right> = -12$. From (\ref{zeromode}) and (\ref{superpotential}) the zero mode is identified as $\psi_0(x)={\rm sech}^2\,x$, whereas the  superpotential, its derivative and the factorization operator are respectively: $\Xi_2(x)=-2\log{\rm cosh} x$, $W_2(x)=2\tanh x$, and $A_2=\frac{d}{dx}+2\tanh \, x$. The supersymmetric partner operator to $K_{2-}=A_2^\dagger A_2$ is also reflectionless and P$\ddot{\rm o}$sch-Teller, such that
\[
K_{2+}=A_2 A_2^\dagger = -\frac{d^2}{dx^2} + 4 -2 \, {\rm sech}^2 \, x = A_1^\dagger A_1 +3 = K_{1-}+3
\]
defines the next level in the supersymmetric hierarchy.
The $K_{1-}$ operator in the formula above in turn factorizes
in the form $K_{1-}=A_1^\dagger A_1$ where $A_1$ is the first-order
operator arising for sine-Gordon kinks. We have implicitly used the
fact that $\omega_1^2=3$ is the first excited eigenvalue of $K_{2-}$
and the ground state energy of $K_{2+}$. The supersymmetric partner
operator to $K_{1-}$ is, as we know from the sine-Gordon kink analysis,
\[
K_{1+}= A_1 A_1^\dagger =-\frac{d^2}{dx^2}+1 \hspace{0.4cm} ,
\]
which fixes $v_1^2=1$. The phase shift associated with this free particle operator is null: $\delta^{(1+)}(k)=0$. Again, the SUSY quantum mechanics hierarchy allows us to apply formula (\ref{gdhn3}):
\[
\frac{\Delta E(\phi_K)}{\hbar \gamma_d^2} = \frac{1}{2} \sqrt{3} - \frac{12}{4\pi} + \frac{1}{2\pi} \int_0^\infty dk
\left[ 0 - \frac{-6+4}{\sqrt{k^2+4}} - \frac{2\sqrt{k^2+4}}{k^2+1} \right] = \frac{1}{2} \sqrt{3} -
\frac{3}{\pi}-\frac{1}{\sqrt{3}}= \frac{1}{2\sqrt{3}} - \frac{3}{\pi} \hspace{0.4cm} .
\]
The one-loop kink mass quantum correction obtained this way is exactly the classical result of
Dashen, Hasslacher and Neveu \cite{Dashen1974,Dashen1975,Dashen1975b}.

\subsubsection{One-loop kink mass correction in the parent potential models}

In references \cite{Christ1975,Trullinger1987} the authors introduced and discussed a hierarchy of interesting (1+1)-dimensional field theory models (which were referred to as parent models) whose action functional (\ref{action}) is specified by the potential terms
\begin{equation}
U^{(N)}(\phi)=\frac{2 \Gamma[\frac{1}{2}+\frac{N}{2}]^2}{\pi \Gamma[\frac{N}{2}]^2} \left(1- I^{-1}\textstyle
\left[|\phi|;\frac{1}{2},\frac{N}{2} \right]\right)^N \hspace{0.4cm} .
\label{potentialinPhi}
\end{equation}
Here, $N\in \mathbb{N}$ is a natural number and $I^{-1}[x;a,b]$ is the inverse of the regularized incomplete beta function (or regularized
beta function) $I(x;a,b)$. The authors also identified the kink solutions in these models
\begin{equation}
\phi_K^{(N)}(x)= {\rm Sign}(x)\,\, I[\textstyle \tanh^2
x;\frac{1}{2},\frac{N}{2}]\hspace{0.4cm} ,
\label{kinkN2}
\end{equation}
 all of them connecting the vacuum points $\phi_V=\pm 1$. The second-order differential operators governing these kink fluctuations are precisely the reflectionless P\"oschl-Teller Schr$\ddot{\rm o}$dinger operators shifted in such a way that the ground state energy is zero:
\begin{equation}
K_{N-}= -\frac{d^2}{dx^2} + N^2- N  (N  +1) \, {\rm sech}^2 x \hspace{0.4cm} .
\label{PoschtTellerPotential}
\end{equation}
In fact the motivation in these works is the identification of the field theory models whose second-order kink fluctuation operator is the Schr\"odinger operator (\ref{PoschtTellerPotential}). These
operators have been studied in depth in the literature. The discrete spectrum of $K_{N-}$ is formed by the discrete set of eigenvalues $\omega_n^2=n(2N-n)$, $n=0,1,\dots,N-1$, together with a half-bound state, $\omega_N^2=N^2$, at the threshold of the continuous spectrum, see e.g. \cite{Mateos2010,Correa2008}. The phase
shifts, however, are defined in terms of very complicated hypergeometric functions. This is the point where we shall
take advantage of (\ref{gdhn4}). This formula computes the kink mass quantum correction in terms of the
phase shift of any operator belonging to the supersymmetric hierarchy that, in these models, ends in the free Helmholtz operator where the phase shifts vanish. The application of formula (\ref{gdhn4}) to obtain the mass shifts due to the kink (\ref{kinkN2}) fluctuations will be thus straightforward.

From (\ref{PoschtTellerPotential}) we read $v_{{}_N}^2=N^2$ and $V_{N-}(x)=- N (N  +1) \, {\rm sech}^2
x$, such that the area enclosed by this function is $\langle V_{N-}(x)\rangle=-2N(N+1)$. The
superpotential and its derivative, in turn, are respectively $\Xi_N(x)=-N\log\cosh x$ and $W_N(x)=N \tanh x$. $K_{N-}=A^\dagger_N A_N$ factorizes in term of the first-order operator: $A_N=\frac{d}{dx}+N \tanh x$. The supersymmetric partner operator is thus:
\begin{equation}
K_{N+}= -\frac{d^2}{dx^2} + (N-1)^2- (N-1)N \, {\rm sech}^2 \,x +2N-1 = K_{(N-1)-}+2N-1 \label{nsush}
\end{equation}
We also recognize in formula (\ref{nsush}) the $K_{(N-1)-}$ operator arising in the next level of the supersymmetric hierarchy and identify $v_{(N-1)}^2=(N-1)^2$ as well as $V_{N+}(x)=V_{(N-1)-}(x)=- (N-1)N \, {\rm sech}^2 x$. The first excited bound state eigenvalue of $K_{N-}$ is $\omega_1^2=v_{{}_N}^2-v_{{}_{(N-1)}}^2=N^2-(N-1)^2=2N-1$, as requested. Applying this process iteratively, the hierarchy of the supersymmetric pairs of operators is constructed. The superpotentials and their derivatives in the $j$-th iteration are: $\Xi_{(N-j)}(x)=-(N-j)\log \cosh x$, $W_{(N-j)}(x)=(N-j) \tanh x$. In the next array we collect the pertinent information
\begin{eqnarray*}
K_{(N-j)-}&=&-\frac{d^2}{dx^2}+(N-j)^2-(N-j)(N-j+1){\rm sech}^2 x \, \, \, , \, \, \, v_{(N-j)}^2=(N-j)^2\\
K_{(N-j)+}&=&-\frac{d^2}{dx^2}+(N-j-1)^2-(N-j-1)(N-j){\rm sech}^2 x +2(N-j)-1=\\ &=& K_{(N-j-1)-}+2(N-j)-1
\end{eqnarray*}
In the last iteration, $j=N-1$, we find
that $K_{1+}=-\frac{d^2}{dx^2}+1$ and therefore $\delta^{(1+)}(k)=0$. Plugging the $j=N-1$ information into equation (\ref{gdhn4}) we obtain
\[
\frac{\Delta E(\phi_K)}{\hbar\gamma_d^2} = \frac{1}{2}\sum_{n=1}^{N-1} \sqrt{n(2N-n)} - \frac{N(N+1)}{2\pi}
+\frac{1}{2\pi} \int_0^\infty dk \left[ \frac{N(N-1)}{\sqrt{k^2 + N^2}} - \sum_{j=1}^{N-1} \frac{2v_{{}_{(N-j)}}\sqrt{k^2 +
N^2}}{k^2 + v_{{}_{(N-j)}}^2} \right] \, \, .
\]
Because $\sum_{j=1}^{N-1}\, (N-j)=2N(N-1)$ we write the previous formula in the form
\begin{eqnarray*}
\frac{\Delta E(\phi_K)}{\hbar\gamma_d^2}&=& \frac{1}{2}\sum_{n=1}^{N-1} \sqrt{n(2N-n)} - \frac{N(N+1)}{2\pi}  +\frac{1}{\pi}
\int_0^\infty dk \sum_{j=1}^{N-1} v_{(N-j)} \left[  \frac{1}{\sqrt{k^2 + N^2}} - \frac{\sqrt{k^2 + N^2}}{k^2 + v_{(N-j)}^2} \right] =
\\&=& \frac{1}{2}\sum_{n=1}^{N-1} \sqrt{n(2N-n)} - \frac{N(N+1)}{2\pi}  -\frac{1}{\pi} \sum_{j=1}^{N-1} \sqrt{j(2N-j)}
\arccos \frac{N-j}{N} =  \\&=&  - \frac{1}{2\pi} N(N+1)+ \frac{1}{\pi} \sum_{j=1}^{N-1} \sqrt{j(2N-j)}  \arcsin
\frac{N-j}{N} \hspace{0.4cm} .
\end{eqnarray*}
The one-loop kink mass correction in the parent models (\ref{potentialinPhi}) is finally
\begin{equation}
\frac{\Delta E(\phi_K^{(N)})}{\hbar\gamma_d^2} = - \frac{1}{2\pi} N(N+1) + \frac{1}{\pi} \sum_{r=1}^{N-1}
\sqrt{N^2-r^2}  \arcsin \frac{r}{N} \hspace{0.4cm} ,
\label{correction}
\end{equation}
where we have redefined $r=N-j$ in the summation index. The same result was found by Boya and Casahorran in
\cite{Boya1989} by applying the procedure of Cahill, Comtet and Glauber \cite{Cahill1976}, which gives the one-loop kink mass correction in terms of the bound state eigenvalues of the $K_{N-}$ operator, see \cite{Boya1989}. We stress that
this last procedure is valid only if the potential well is reflectionless.

\section{Exotic scalar field theory models with exotic kinks}

In this section we shall discuss other scalar field theory models, where the issue of the computation of the one-loop correction to the kink masses is an analytical nightmare if one uses the standard approaches to the problem. We shall show, however, that in these exotic cases the supersymmetric quantum mechanical techniques works well in computations of the kink mass shifts.

\subsection{Kink mass quantum shift in the $U(\phi)= \frac{1}{2} \phi^2 \cos^2 \log |\phi|$ model }

The action (\ref{action}) in this model is specified by the potential energy density
\[
U(\phi)= \frac{1}{2} \phi^2 \cos^2 \log |\phi| \hspace{0.4cm} .
\]
This function is depicted in Figure 3. Notice that $U(\phi)$ as a function of $\phi$ oscillates indefinitely around the origin. We check, however, that $U(\phi)\in {\cal C}^\infty(\mathbb{R}-\{0\})$, although only $U(\phi)\in {\cal C}^1(\mathbb{R})$ in the whole real line. This model was addressed in reference \cite{Kumar2003}, where the authors succeeded in identifying the kink type solutions.

\begin{figure}[ht]
\centerline{\includegraphics[height=3cm]{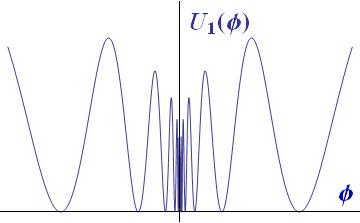}\hspace{1cm}
\includegraphics[height=3cm]{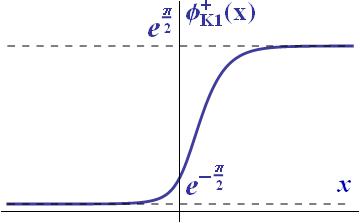} }

\caption{\small Re-scaled graphics of $U(\phi)$ (left) and plot of the kink $\phi_{K1}^+$ connecting the vacua $\phi_{V_+}^{(-1)}$ and $\phi_{V_+}^{(0)}$ (right).}
\end{figure}

The set of vacua ${\cal M}$ of the model is the set of zeroes of $U(\phi)$: $\phi_V^{(j)}=0 \, \equiv \, \log |\phi_V^{(j)}|=\pi(j+\frac{1}{2})$, $j\in\mathbb{Z}$. Therefore, the set ${\cal M}$ is the union of two subsets separated by the origin which is a singular point:
\[
{\cal M} = {\cal M}_+ \cup {\cal M}_- \, , \hspace{0.5cm}
{\cal M}_+= \{ \phi_{V_+}^{(j)}\} = e^{\pi(j+\frac{1}{2})}\, , \hspace{0.5cm}
{\cal M}_-= \{ \phi_{V_-}^{(j)}\} =- e^{\pi(j+\frac{1}{2})} ,\hspace{0.5cm} .
\]
Thus, there is an infinite number of vacua in this model:
\[
\frac{\partial^2 U}{\partial\phi^2}=-\frac{1}{2}(1+\cos(\log \phi^2)-\sin(\log \phi^2)(3+|\phi|\frac{\partial^2\vert\phi\vert}{\partial\phi^2})) \, \, \, , \hspace{0.3cm} \frac{\partial^2 U}{\partial\phi^2}\left. \right|_{\phi_{V_+}^{(j)}}=\frac{\partial^2 U}{\partial\phi^2}\left. \right|_{\phi_{V_-}^{(j)}}=1 \, \, , \, \forall j \, \, ,
\]
all of them being equivalent in the sense that the masses of the fundamental quanta emerging from any vacuum are: $v_1^2=1$. The distance in field space between consecutive vacua, however, differs with $j$: $d(\phi_{V_\pm}^{(j+1)},\phi_{V_\pm}^{(j)})= \left|\phi_{V_\pm}^{(j+1)}-\phi_{V_\pm}^{(j)}\right|=e^{\frac{\pi}{2}+ \pi j}(e^\pi-1)$.

The kink solutions of the first-order equation (\ref{ode1}) are
\begin{equation}
\phi_{K+}^{(j)}(x)=e^{{\rm gd}(x)+ \pi j}=e^{2\arctan \tanh \frac{x}{2}+\pi j} , \hspace{0.4cm}
\phi_{K-}^{(j)}(x)=-e^{2\arctan \tanh \frac{x}{2}+\pi j} \hspace{0.4cm} ,
\label{kink1}
\end{equation}
which respectively interpolate between the $\phi_{V\pm}^{(j)}$ and $\phi_{V\pm}^{(j+1)}$ vacua {\footnote{
${\rm gd}(x)$ is the standard notation for the Gudermannian.}}. In Figure 3, a particular kink connecting the $\phi_{V+}^{(-1)}$ and the $\phi_{V+}^{(0)}$ vacua is depicted.

In this section we shall compute the one-loop correction to the classical mass $E[\phi_{K\pm}^{(j)}]=\frac{2}{5}e^{2\pi j}\cosh\pi$ of these
kink solutions (\ref{kink1}). The second-order operator governing these kink fluctuations is
\[
K_{1-}=-\frac{d^2}{dx^2}+1-{\rm sech}^2 x-3 \, {\rm sech}\,x \tanh x \hspace{0.4cm} .
\]
This sets $V_{1-}(x)=-{\rm sech}^2 x-3 \, {\rm sech}\,x \tanh x$, which is the Scarf II potential, and therefore
$\left<V_{1-}(x)\right> = -2$. The zero modes are identified from condition (\ref{zeromode})
\[
\psi_0^{(j)}(x)={\rm sech}\, x \, e^{2\arctan \tanh \frac{x}{2}+ \pi j}
\]
and by using (\ref{superpotential}) the derivative of the superpotential is $W_1(x)=\tanh x - \, {\rm
sech}\,x$. This is a remarkable fact: despite being of different \lq\lq length \rq\rq the kinks labeled by different $j$ give rise to identical wells{\footnote{The connection of the (inverse) Gudermannian to the Mercator projection is well known. Lengths are distorted in such a way that different parallels look similar: the parallels closer to the Poles grow longer. The kinks in this model exhibit a similar phenomenon but the rescaling is exponential.}}. The second-order kink fluctuation operator $K_{1-}$ factorizes in the form
\[
K_{1-}= A_1^\dagger A_1 \hspace{0.5cm} \mbox{with} \hspace{0.5cm}  A_1=\frac{d}{dx} + \tanh x - \, {\rm
sech}\,x \hspace{0.5cm} \mbox{and} \hspace{0.5cm}
A_1^\dagger = -\frac{d}{dx} + \tanh x - \, {\rm sech}\,x \hspace{0.4cm} ,
\]
and consequently the supersymmetric partner operator to $K_{1-}$ reads:
\[
K_{1+}=A_1A_1^\dagger=-\frac{d^2}{dx^2} +1 + \, {\rm sech}^2 \, x - \, {\rm sech} \, x \tanh x
\hspace{0.4cm} .
\]
Although the $K_{1-}$ and $K_{1+}$ operators look similar, the analytical expressions of the eigenfunctions in the continuous spectrum of $K_{1+}$ are much simpler than those associated with the $K_{1-}$ operator. This allows us to extract the spectral data, the bound state energies and the phase shifts of this operator, in an easier form and reduce the calculations required to obtain the kink mass quantum shift. Figure 4 shows the potential wells arising respectively in the $K_{1-}$ and $K_{1+}$ operators.

\begin{figure}[ht]
\centerline{\includegraphics[height=3cm]{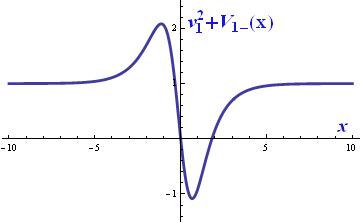} \hspace{2cm}
\includegraphics[height=3cm]{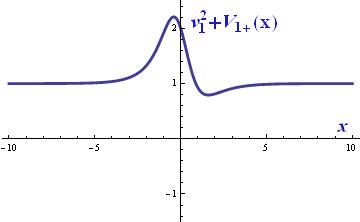}}

\caption{\small Potential wells of the $K_{1-}$ (left) and $K_{1+}$ operators (right).}
\end{figure}

\noindent The eigenfunctions in the continuous spectrum of $K_{1+}$ are
\begin{eqnarray*}
\psi_1(x)&=& e^{\arctan \sinh x} \, {}_2F_1 \Big[-ik,ik,\frac{1}{2}-i,\frac{1+i\sinh x}{2}\Big] \hspace{0.4cm}, \\
\psi_2(x)&=&  \Big(\frac{1+i\sinh x}{2}\Big)^{\frac{1}{2}+i} e^{\arctan \sinh x} \,
{}_2F_1\Big[\frac{1}{2}+i-ik,\frac{1}{2}+i+ik,\frac{3}{2}+i, \frac{1+i\sinh x}{2}\Big]\hspace{0.4cm},
\end{eqnarray*}
where $\omega^2=k^2+1$. The asymptotic behavior of these two linearly independent eigenfunctions for $x\rightarrow
-\infty$ is
\begin{eqnarray*}
\psi_1(x) &\stackrel{x\rightarrow -\infty}{\longrightarrow} & e^{-\frac{\pi}{2}(k+1)}
\frac{\Gamma[\frac{1}{2}-i]\Gamma[\frac{1}{2}+ik]}{2\sqrt{\pi} \Gamma[\frac{1}{2}-i+ik]} e^{-ikx} +
e^{-\frac{\pi}{2}(-k+1)} \frac{\Gamma[\frac{1}{2}-i]\Gamma[\frac{1}{2}-ik]}{2\sqrt{\pi} \Gamma[\frac{1}{2}-i-ik]}
e^{ikx} \hspace{0.4cm},\\
\psi_2(x) &\stackrel{x\rightarrow - \infty}{\longrightarrow} &  -e^{-\frac{\pi}{2}(k-1)}
\frac{\Gamma[\frac{1}{2}+ik]\Gamma[\frac{3}{2}+i]}{2\sqrt{\pi}k \Gamma[\frac{1}{2}+i+ik]} e^{-ikx} +
e^{\frac{\pi}{2}(k+1)} \frac{\Gamma[\frac{1}{2}-ik]\Gamma[\frac{3}{2}+i]}{2\sqrt{\pi}k \Gamma[\frac{1}{2}+i-ik]}
e^{ikx}\hspace{0.4cm},
\end{eqnarray*}
whereas for $x\rightarrow +\infty$ it becomes
\begin{eqnarray*}
\psi_1(x) &\stackrel{x\rightarrow \infty}{\longrightarrow} & e^{\frac{\pi}{2}(k+1)}
\frac{\Gamma[\frac{1}{2}-i]\Gamma[\frac{1}{2}+ik]}{2\sqrt{\pi} \Gamma[\frac{1}{2}-i+ik]} e^{ikx} +
e^{\frac{\pi}{2}(-k+1)} \frac{\Gamma[\frac{1}{2}-i]\Gamma[\frac{1}{2}-ik]}{2\sqrt{\pi} \Gamma[\frac{1}{2}-i-ik]}
e^{-ikx}\hspace{0.4cm},\\
\psi_2(x) &\stackrel{x\rightarrow \infty}{\longrightarrow}& e^{\frac{\pi}{2}(k-1)}
\frac{\Gamma[\frac{1}{2}+ik]\Gamma[\frac{3}{2}+i]}{2\sqrt{\pi}k \Gamma[\frac{1}{2}+i+ik]} e^{ikx} -
e^{-\frac{\pi}{2}(k+1)} \frac{\Gamma[\frac{1}{2}-ik]\Gamma[\frac{3}{2}+i]}{2\sqrt{\pi}k \Gamma[\frac{1}{2}+i-ik]}
e^{-ikx} \hspace{0.4cm} .
\end{eqnarray*}
A linear combination of the continuous functions $\psi_1(x)$ and $\psi_2(x)$ describes an incident wave
coming from the left side, behaving asymptotically as:
\begin{eqnarray*}
e^{ikx} + \frac{\pi e^{-k\pi}\Gamma[\frac{1}{2}+ik]\left( {\rm sech}\,(k+1)\pi -e^{2\pi} \, {\rm sech}\,(1-k)\pi
\right)}{(1+e^{2\pi})\Gamma[\frac{1}{2}-ik] \Gamma[\frac{1}{2}-i+ik]\Gamma[\frac{1}{2}+i+ik]} e^{-ikx} & \hspace{1cm} &
x\rightarrow -\infty \\
\frac{\pi e^{\pi}\Gamma[\frac{1}{2}+ik]\left( {\rm sech}\,(k+1)\pi + \, {\rm sech}\,(1-k)\pi \right)}{(1
+e^{2\pi})\Gamma[\frac{1}{2}-ik] \Gamma[\frac{1}{2}-i+ik]\Gamma[\frac{1}{2}+i+ik]} e^{ikx} &\hspace{1cm} & x\rightarrow
+\infty
\end{eqnarray*}
This result determines the reflection and transmission scattering amplitudes of this incident wave:
\begin{eqnarray*}
\rho_{1+}^{(\rightarrow)}(k) &=&\frac{\pi e^{-k\pi}\Gamma[\frac{1}{2}+ik]\left( {\rm sech}\,(k+1)\pi -e^{2\pi} \, {\rm
sech}\,(1-k)\pi \right)}{(1+e^{2\pi})\Gamma[\frac{1}{2}-ik] \Gamma[\frac{1}{2}-i+ik]\Gamma[\frac{1}{2}+i+ik]}
\hspace{0.4cm},\\
\sigma_{1+}^{(\rightarrow)}(k) &=& \frac{\pi e^{\pi}\Gamma[\frac{1}{2}+ik]\left( {\rm sech}\,(k+1)\pi + \, {\rm
sech}\,(1-k)\pi \right)}{(1 +e^{2\pi})\Gamma[\frac{1}{2}-ik] \Gamma[\frac{1}{2}-i+ik]\Gamma[\frac{1}{2}+i+ik]}
\hspace{0.4cm} .
\end{eqnarray*}
In a similar way, the asymptotic behavior of an incident wave coming from the right side is
\begin{eqnarray*}
\frac{\pi e^{\pi}\Gamma[\frac{1}{2}+ik]\left( {\rm sech}\,(k+1)\pi + \, {\rm sech}\,(1-k)\pi \right)}{(1
+e^{2\pi})\Gamma[\frac{1}{2}-ik] \Gamma[\frac{1}{2}-i+ik]\Gamma[\frac{1}{2}+i+ik]} e^{-ikx} & \hspace{1cm} &
x\rightarrow -\infty \\
e^{-ikx} + \frac{\pi e^{k\pi}\Gamma[\frac{1}{2}+ik]\left( e^{2\pi}{\rm sech}\,(k+1)\pi -\, {\rm sech}\,(1-k)\pi
\right)}{(1+e^{2\pi})\Gamma[\frac{1}{2}-ik] \Gamma[\frac{1}{2}-i+ik]\Gamma[\frac{1}{2}+i+ik]} e^{ikx} & \hspace{1cm} &
x\rightarrow +\infty \, \, ,
\end{eqnarray*}
such that the reflection and transmission scattering amplitudes are:
\begin{eqnarray*}
\rho_{1+}^{(\leftarrow)}(k) &=&\frac{\pi e^{k\pi}\Gamma[\frac{1}{2}+ik]\left( e^{2\pi}{\rm sech}\,(k+1)\pi -\, {\rm
sech}\,(1-k)\pi \right)}{(1+e^{2\pi})\Gamma[\frac{1}{2}-ik] \Gamma[\frac{1}{2}-i+ik]\Gamma[\frac{1}{2}+i+ik]}
\hspace{0.4cm},\\
\sigma_{1+}^{(\leftarrow)}(k) &=&\frac{\pi e^{\pi}\Gamma[\frac{1}{2}+ik]\left( {\rm sech}\,(k+1)\pi + \, {\rm
sech}\,(1-k)\pi \right)}{(1 +e^{2\pi})\Gamma[\frac{1}{2}-ik] \Gamma[\frac{1}{2}-i+ik]\Gamma[\frac{1}{2}+i+ik]}
\hspace{0.4cm} .
\end{eqnarray*}
Note that $\sigma_{1+}^{(\rightarrow)}(k)=\sigma_{1+}^{(\leftarrow)}(k)=\sigma_{1+}(k)$ and
$\rho_{1+}^{(\leftarrow)}(k)=-\rho_{1+}^{(\rightarrow)}(k)$. The unitary scattering matrix follows:
\[
S_{1+}=\left( \begin{array}{cc} \sigma_{1+} & \rho_{1+}^{(\leftarrow)}\\ - \rho_{1+}^{(\leftarrow)} & \sigma_{1+}
\end{array} \right) \hspace{0.4cm} .
\]
The phase shifts are the arguments of the modulus-one eigenvalues of this matrix
\[
e^{2i\delta_\pm^{({1+})}} =\sigma_{1+} \pm i \rho_{1+}^{(\leftarrow)}= \frac{\pi
\Gamma[\frac{1}{2}+ik]}{\Gamma[\frac{1}{2}-ik] \Gamma[\frac{1}{2}-i+ik] \Gamma[\frac{1}{2}+i+ik] (\sinh \pi \pm i\cosh
k\pi)} \hspace{0.4cm} .
\]
The derivative of the total phase shift $\delta^{(1+)}(k) =\delta^{(1+)}_+(k)+\delta^{(1+)}_-(k)$ with respect to the momentum $k$ becomes
\begin{equation}
\frac{d \delta^{({1+})}(k)}{dk} = {\rm Re}\left[ \psi({\textstyle\frac{1}{2}}-ik) + \psi({\textstyle\frac{1}{2}}+ik) -
\psi({\textstyle\frac{1}{2}}-i+ik) - \psi({\textstyle\frac{1}{2}}+i+ik) \right] \hspace{0.4cm} ,
\label{phase2}
\end{equation}
where $\psi(z)$ is the digamma function. We stress here that the calculation of the spectral density coming from the  $K_{1+}$
scattering waves is much easier to achieve than if we had handled the corresponding computation for the $K_{1-}$ operator. Note
that $K_{1-}$ has only one bound state, the zero mode, because there are no bound states in the spectrum of $K_{1+}$. Plugging
(\ref{phase2}) into formula (\ref{gdhn3}), the one-loop correction to the (\ref{kink1}) kink
masses can be figured out:
\[
\frac{\Delta E(\phi_K)}{\hbar \gamma_d^2} = -\frac{1}{4} + \frac{1}{4\pi} \cdot (-2) + \frac{1}{2\pi} \int_0^\infty
dk\left[ \frac{d \delta^{({1+})}(k)}{dk} \sqrt{k^2+1} - \frac{1}{\sqrt{k^2+1}} \right] \hspace{0.4cm} .
\]
This expression requires the integration of a complicated function, which we estimate numerically to find:
$\int_0^\infty dk [ \frac{d \delta^{({1+})}(k)}{dk} \sqrt{k^2+1}-\frac{1}{\sqrt{k^2+1}} ]\approx -1.1823055$.
Therefore, the final response to the mass quantum shift  of all the kinks in this model is
\[
\frac{\Delta E(\phi_K)}{\hbar \gamma_d^2} \approx -0.597325 \hspace{0.4cm} .
\]

\subsection{A model reconstructed from a zero mode}

The procedure proposed in reference \cite{Christ1975} shows how to construct a field theory model from a specific function selected as the zero-mode of a conjectural second-order kink fluctuation operator. This procedure has been used by Kumar in order to generate models with very general soliton profiles, see \cite{Kumar1987,Dey1994}. The steps to be followed in order to implement this idea are described at the end of section 2.1. Here, we start from a particular selection of the zero-mode (see Figure 5, left):
\[
\psi_0(x)= e^{-{\rm gd}(x)}\, {\rm sech}^\frac{1}{2} x=e^{-2\arctan \tanh \frac{x}{2}} \, {\rm sech}^\frac{1}{2} x \hspace{0.4cm},
\]
This function is the ground state of the following second-order fluctuation operator
\begin{equation}
K_{1-}=-\frac{d^2}{dx^2} + \frac{1}{4} + \frac{1}{4} \, {\rm sech}^2 x + 2 \, {\rm sech}\,x \tanh
x \hspace{0.4cm}.
\label{Hessian03}
\end{equation}

\begin{figure}[ht]
\centerline{\includegraphics[height=3cm]{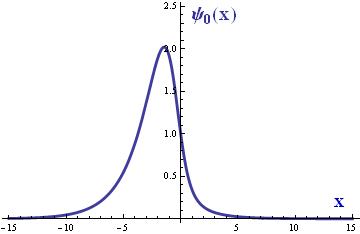}\hspace{1cm}
\includegraphics[height=3cm]{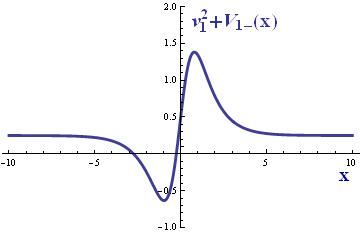}\hspace{1cm}
\includegraphics[height=3cm]{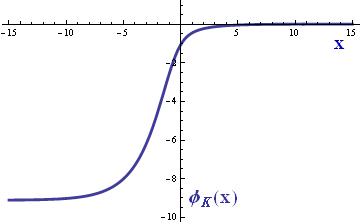} }

\caption{\small Zero mode (left), potential well (middle) and kink (right) encompassed in the second-order kink
fluctuation operator $K_{1-}$.}
\end{figure}

The kink profile is identified through the integration of the zero mode, (\ref{kinkfromzeromode}):
\[
\phi_K(x)=\int \psi_0(x) dx = \int e^{-2\arctan \tanh \frac{x}{2}} \, {\rm sech}^\frac{1}{2} x \,dx = \int e^{-z}
\sqrt{\sec z} \, dz \hspace{0.4cm}.
\]
Here, the change of variable, $z=\xi(x)$, where $\xi:\mathbb{R} \rightarrow (-\frac{\pi}{2},\frac{\pi}{2})$, $x\mapsto
\xi(x)= 2 \arctan \tanh \frac{x}{2}$, has been performed for the sake of simplicity. The inverse of this diffeomorphism
is $x= \xi^{-1}(z)=2 \, {\rm arctanh}\, \tan \frac{z}{2}$ and, in particular, $dx=\sec z dz$. We also define the
transcendent function $\Upsilon:[-\frac{\pi}{2},\frac{\pi}{2}] \rightarrow \mathbb{R}$ as $\Upsilon(z)=\int_0^z e^{-t}
\sqrt{\sec t}\,dt$ such that the kink solution
\begin{equation}
\phi_K(x)=\Upsilon(2 \arctan \tanh {\textstyle\frac{x}{2}})
\label{Kink03}
\end{equation}
is expressed in terms of the hypergeometric function in the form
\[
\phi_K(x)= {\textstyle {\rm Re} \left\{ (\frac{8}{5} - \frac{4 i }{5} ) e^{-2(1+i) \arctan \tanh \frac{x}{2}} \,
\sqrt{{\rm sech}\, x} \left(
-1+{}_2F_1\left[-\frac{1}{4}+\frac{i}{2},1,\frac{1}{4}+\frac{i}{2},\frac{(e^x-i)^2}{(e^x+i)^2}\right] \right) \right\}}
\hspace{0.4cm}.
\]
The classical kink energy, however, is easy to compute:
\[
E[\phi_k]= \int_{-\infty}^\infty \, \psi_0^2(x)\, dx = \int_{-\infty}^\infty \, {\rm exp}[-2 {\rm gd}(x)]{\rm sech}x\, dx= {\rm sinh}\, \pi \, \, .
\]
Finally, the potential energy density  $U(\phi)$ that characterizes the scalar field theory model is determined from
(\ref{potentialfromzeromode}) in the kink profile range:
\[
U(\phi) = \frac{1}{2} e^{2 \Upsilon^{-1}(\phi)} \cos [\Upsilon^{-1}(\phi)] \hspace{0.4cm}.
\]
The second-order kink fluctuation operator (\ref{Hessian03}) sets the values $v_1^2=\frac{1}{4}$ and $V_{1-}(x)=
\frac{1}{4} \, {\rm sech}^2 x + 2 \, {\rm sech}\, x \tanh x$. Hence $\left< V_{1-}(x) \right>=\frac{1}{2}$ holds and
the function $W_1(x)$ allowing the factorization of the $K_{1-}$ operator is $W_1(x)=-\frac{d}{dx}\log\psi_0(x)=\frac{1}{2} \tanh x + {\rm
sech}\, x$. In order to estimate the mass quantum shift we need to extract the spectral information of the
operator (\ref{Hessian03}). This task is arduous. We shall employ (\ref{gdhn2}) to circumvent this difficulty. This
formula expresses the mass shift in terms of the spectral properties of the supersymmetric partner operator $K_{1+}$,
\begin{equation}
K_{1+}=-\frac{d^2}{dx^2} + \frac{1}{4} + \frac{5}{4} \, {\rm sech}^2 x \hspace{0.4cm}.
\label{Hessian05}
\end{equation}
Fortunately, (\ref{Hessian05}) is a differential operator with a P\"oschl-Teller type potential well whose spectral
properties are well known. In particular, the spectrum of $K_{1+}$ lacks bound states such that $K_{1-}$ presents only one bound state, the zero mode. The spatial reflection symmetry of the problem permits us to obtain the phase shifts needed in (\ref{gdhn2}) only from the continuous spectrum eigenfunctions:
\begin{equation}
\psi_k(x)= C P_{-\frac{1}{2}+i}^{ik}(\tanh x) \hspace{0.4cm}.
\label{continuous04}
\end{equation}
Here, $P_\mu^\nu(z)$ stands for Legendre polynomials and $C$ is an integration constant.

Choosing $C=\frac{\Gamma[\frac{1}{2}-i(1+k)] \Gamma[\frac{1}{2}+i(1-k)]}{\Gamma[-ik]}$ the asymptotic behavior of the
eigenfunctions (\ref{continuous04}) for $x\rightarrow -\infty$ is:
\[
\psi_k(x) \stackrel{x\rightarrow -\infty}{\longrightarrow}   e^{ikx} +
\frac{\cosh (\pi) \Gamma[ik]\Gamma[\frac{1}{2}-i(1+k)] \Gamma[\frac{1}{2}+i(1-k)]}{\pi\Gamma[-ik]}
e^{-ikx} \hspace{0.4cm},
\]
whereas for $x\rightarrow \infty$ it becomes
\[
\psi_k(x) \stackrel{x\rightarrow \infty}{\longrightarrow}
\frac{\Gamma[\frac{1}{2}-i(1+k)] \Gamma[\frac{1}{2}+i(1-k)]}{\Gamma[-ik]\Gamma[1-ik]}
e^{ikx}\hspace{0.4cm}.
\]
Consequently the eigenfunctions (\ref{continuous04}) describe incident waves coming from the left. The $K_{1+}$ reflection and transmission scattering amplitudes read
\[
\rho_{1+}^{(\rightarrow)}=\frac{\cosh (\pi) \Gamma[ik]\Gamma[\frac{1}{2}-i(1+k)]
\Gamma[\frac{1}{2}+i(1-k)]}{\pi\Gamma[-ik]}  \hspace{0.5cm},\hspace{0.5cm}
\sigma_{1+}^{(\rightarrow)} = \frac{\Gamma[\frac{1}{2}-i(1+k)] \Gamma[\frac{1}{2}+i(1-k)]}{\Gamma[-ik]\Gamma[1-ik]}
\hspace{0.4cm}.
\]
Moreover, $\sigma_{1+}^{(\rightarrow)}=\sigma_{1+}^{(\leftarrow)}=\sigma_{1+}$ and
$\rho_{1+}^{(\rightarrow)}=\rho_{1+}^{(\leftarrow)}=\rho_{1+}$ hold because of the symmetry of $K_{1+}$. The
eigenvalues $\sigma_{1+}\pm \rho_{1+}=e^{2i\delta^{({1+})}_\pm}$ of the $S$-matrix provide the phase shifts
$\delta^{({1+})}(k)=\delta^{({1+})}_+(k)+\delta^{({1+})}_-(k)$ such that their derivatives give the spectral density:
\begin{equation}
\frac{d\delta^{({1+})}(k)}{dk} = {\rm Re} \left[{\textstyle 2 \psi(-i k)-\psi(\frac{1}{2}-i(1+k)) -
\psi(\frac{1}{2}+i(1-k))   } \right] \hspace{0.4cm}. \label{phase3}
\end{equation}
Plugging (\ref{phase3}) into formula (\ref{gdhn3}), the one-loop mass quantum correction to this kink
(\ref{Kink03}) can be figured out:
\[
\frac{\Delta E(\phi_K)}{\hbar \gamma_d^2} = -\frac{1}{4} \cdot \frac{1}{2} + \frac{1}{4\pi} \cdot \frac{1}{2} +
\frac{1}{2\pi} \int_0^\infty dk\left[ \frac{d \delta^{({1+})}(k)}{dk} \sqrt{k^2+{\textstyle\frac{1}{4}}} -
\frac{\frac{1}{2}\cdot \frac{1}{2} + 2 \cdot \frac{1}{2}}{\sqrt{k^2+\frac{1}{4}}} \right] \hspace{0.4cm}.
\]
The numerical evaluation of the integral in the previous formula gives: $\frac{1}{2\pi} \int_0^\infty dk\Big[ \frac{d
\delta^{({1+})}(k)}{dk} \sqrt{k^2+\frac{1}{4}} - \frac{\frac{5}{4}}{\sqrt{k^2+\frac{1}{4}}} \Big]\approx -0.152774$.
Therefore, the final response to the kink mass quantum shift in this model is obtained:
\[
\frac{\Delta E(\phi_K)}{\hbar \gamma_d^2} \approx -0.237985\hspace{0.4cm}.
\]

\section{Construction of scalar field model families sharing the kink mass shift}

In references \cite{Kumar1987,Dey1994} Kumar built one-parametric families of (1+1) dimensional scalar
field theory models whose kinks give rise to the same second-order kink fluctuation operator $K_{1-}$. Evidently, the mass quantum shift of all
these kinks will be the same, see (\ref{gdhn}). In this section we shall go one step farther. We envisage the identification of one-parametric families of field theory models whose second-order kink fluctuation operators $K_{1-}$ are different but share the same supersymmetric partner $K_{1+}$. From (\ref{gdhn2}) it is possible to compute the common one-loop mass correction to all these kinks in an easy way. We start from a specific operator $K_{1+}$, which we assume to be factorized in the usual form:
\[
K_{1+}=A_1A_1^\dagger \hspace{1cm} \mbox{where} \hspace{1cm} A_1=\frac{d}{dx}+W_1(x) \hspace{0.4cm}.
\]
The natural question to ask is: does there exist a one-parametric deformation of a superpotential $\int^x W_1(z) dz$
supplying the same factorization as the reverse $K_{1+}$ factorization? The answer is provided by a family of the superpotentials such that:
\begin{equation}
W_1(x,C) = W_1(x) + \frac{e^{2 \Xi_1(x)}}{C+\int^x
e^{2\Xi_1(\eta)}d\eta}\,, \qquad \Xi_1(x)=-\int^x \, W_1(\xi)d\xi
\,.
\label{GenericSuperpotential+}
\end{equation}
We shall restrict the real parameter $C$ to a range
 such that $W_1(x,C)$ is non-singular.
With this definition we have a natural relation
$W_1(x,\pm\infty) = W_1(x)$, and the zero modes of the family of operators
\begin{equation}\label{K1-Cadd}
K_{1-}(C)=A_1^\dagger(C) A_1(C)=-\frac{d^2}{dx^2} +  W_1(x,C)^2 -
W_1^\prime(x,C) \, \, , \hspace{0.5cm} \mbox{where} \hspace{0.5cm}
A_1(C)=\frac{d}{dx}+W_1(x,C) \, \, ,
\end{equation}
are: $\psi_0(x,C)=e^{-\int^x W_1(\xi,C)d\xi}$.

One easily checks that for any $W_1(x,C)$ of the  form
(\ref{GenericSuperpotential+}) there is a SUSY partner,
$K_{1+}$, common to all the members in the $K_{1-}(C)$ family:
\[
K_{1+}=A_1(C) A_1^\dagger(C)= A_1 A_1^\dagger \, \, , \, \, \,
\forall C \hspace{0.4cm}.
\]
This can be understood from the results of \cite{Plyushchay2012}. Taking into account the
relations $A_1(C)K_{1-}(C)=K_{1+}A_1(C)$ and $K_{1-}(C)A_1^\dagger(C)=A_1^\dagger(C)K_{1+}$, one can immediately say that within such a family, any two Schr$\ddot{\rm o}$dinger operators are related by the second-order supersymmetry  generated by the second-order intertwining generators,
 $Y(C,C^\prime)K_{1-}(C^\prime)=K_{1-}(C)Y(C,C^\prime)$, where
\begin{equation}\label{YCC-add}
 Y(C,C^\prime)=A_1^\dagger(C)A_1(C^\prime).
\end{equation}

The idea now is to reconstruct a scalar field model family from this
$C$-parametric family of Schr$\ddot{\rm o}$dinger operators. The
zero modes are known and consequently the kink solutions of each
member of the family are obtained through spatial integration:
$\phi_K(x,C)=\int^x \,\psi_0(\xi,C) \, d\xi$. Finally, taking into
account that $\phi=\phi_K(x,C)$ are injective functions, they can be
inverted, $x=\phi_K^{-1}(\phi)$, in order to derive the potential
$U(\phi)$ for each family member in the range of the kink profiles:
$U(\phi)=\frac{1}{2} \psi_0^2[\phi_K^{-1}(\phi)]$. We remark that the
particle masses in all the family members are
equal because their kinks fluctuate according to operators with the
same SUSY partners.

\subsection{A family built from the sine-Gordon model}

The superpotential allowing the factorization  of the second-order
sG-soliton fluctuation operator (\ref{solitonhessian}) is
$\Xi_1(x)=-\log\cosh x$. Thus, the family of superpotentials, all of which share
the free particle Hamiltonian as the SUSY partner is:
\begin{equation}
\Xi_1(x,C)= -\log(C \cosh x+\sinh x) \quad , \quad W_1(x,C)
=\frac{{\rm sech}^2 x}{C+\tanh x} +\tanh x \hspace{0.4cm},
\label{superpotential55}
\end{equation}
where $C$ is assumed to lead to non-singular $W_1(x,C)$.  The
second-order soliton fluctuation operator (\ref{Hessian00}) arising
in each member of the family and their common supersymmetric partner
(\ref{partneroperator}) are respectively
\begin{equation}
K_{1-}= -\frac{d^2}{dx^2}+1+\frac{2(1-C^2)}{(\sinh x+C  \cosh x)^2}
\hspace{1cm} , \hspace{1cm} K_{1+}= - \frac{d^2}{dx^2}+1
\hspace{0.4cm}. \label{operator55}
\end{equation}
Thus, we have $v_1^2=1$, $V_{1-}(x,C)=\frac{2(1-C^2)}{( \sinh x+C
\cosh x)^2}$, $V_{1+}(x)=0$ and for later use we list:
$\left<V_{1-}(x,C)\right>=-4$, $\delta^{({1+})}(k)=0$. From
(\ref{superpotential}) and (\ref{kinkfromzeromode}) we immediately
write the zero mode and the $C$-dependent kink:
\[
\psi_0(x,C)= \frac{1}{C\cosh x + \sinh x} \hspace{0.7cm},\hspace{0.7cm}
\phi_K(x,C) = \frac{2}{\sqrt{C^2-1}} \arctan \frac{C\tanh \frac{x}{2}+
1}{\sqrt{C^2-1}} \hspace{0.4cm},
\]
where the restriction $C^2>1$ must be assumed in order  to deal with
real kinks.  In fact, if $C^2<1$  $V_{1-}(x,C)$ would be singular.
Finally, from (\ref{potentialfromzeromode}) we also identify the
family of potentials
\begin{equation}
U(\phi,C)= \frac{1}{2C^2(C^2-1)^2} \left[ (C^2-1) \cos  \left(
\sqrt{C^2-1} \phi \right)  \sqrt{C^2-1} \sin \left( \sqrt{C^2-1}
\phi \right) \right]^2 \hspace{0.4cm}. \label{generalsG}
\end{equation}
The absolute minima of this function
\[
\phi_V^{(j)}=\frac{1}{\sqrt{C^2-1}}\left\{{\rm arctan}
\left[\sqrt{C^2-1}\right]+\pi j\right\} \, \, \, , \, \, \,
j\in\mathbb{Z}
\]
are equivalent in the sense that all of them have the same
curvature: $v_1^2=\frac{\delta^2
U}{\delta\phi^2}\left.\right|_{\phi_V^{(j)}}=1$. As predicted, the
particle masses in the whole family are the same. The classical kink
energy, however, differs, $E[\phi_K(x,C)]=\frac{2}{C^2-1}$, but the
computation of the one-loop kink mass shifts by means of formula
(\ref{gdhn2}) confirms that
\[
\frac{\Delta E[\phi_K(C)]}{\hbar\gamma_d^2}=  \frac{1}{4\pi}  (-4) +
\frac{1}{2\pi} \int_0^\infty dq \left[ - \frac{\frac{1}{2}  (-4)+
2}{\sqrt{k^2 + 1}}\right]=-\frac{1}{\pi}
\]
is $C$-independent and equal to the value found in the  sine-Gordon
model. Why is this so is revealed by consecutively performing a
translation and a dilatation in field space. First, the translation
\[
\xi = \phi + \frac{1
}{\sqrt{C^2-1}} \arctan \sqrt{C^2-1}
\]
produces the following modification of the potential:
\[
U(\xi,C)= \frac{1}{2(C^2-1)} \sin^2 [\sqrt{C^2-1}\, \, \xi] \, \, \, .
\]
Accordingly, the field dilatation  $\Phi=2\sqrt{C^2-1}\xi$ shows that the
family of potentials becomes
\begin{equation}
U(\Phi,C) = \frac{1}{4(C^2-1)} (1-\cos \Phi) \, \, \, \label{fsG} .
\end{equation}
All of them incorporate the sine-Gordon potential times a
$C$-dependent  scale factor. The effect of this scale is to deliver
the correct $C$-dependent kink energy by multiplication with the
sine-Gordon kink energy which is $8$. This statement is tantamount
to saying that the scale fixes the norm of the zero mode because the
kink energy density is precisely the square of the zero-mode wave
function.

The requirement $C^2>1$ suggests the following reparametrization:
$\coth \tau=C$. In terms of the new parameter the  superpotential  (\ref{superpotential55}) and
the function $W_1(x,C)$ become
\[
\Xi_1(x,\tau)= -\log\frac{1}{{\rm sinh}\tau} \cosh(x+\tau)  \quad ,
\quad  W_1(x,\tau)=\tanh (x+ \tau) \hspace{0.4cm},
\]
 showing that the parameter merely gives rise to a translation
  in the spatial coordinate of the proper sine-Gordon  counterparts
  (modulo a $\tau$-dependent integration constant in the case
of the superpotential $\Xi(x,\tau)$): $\Xi(x,0)$, $W_1(x,0)$.  The
kink fluctuation operators $K_{1-}(\tau)=-\frac{d^2}{dx^2}
+1-\frac{2}{\cosh^2(x+\tau)}$ are manifestly isospectral, the
eigenvalues  of any $K_{1-}(\tau)$ operator are clearly independent
of the $\tau$ parameter. The reconstruction of the field theory
model from the zero mode, $\psi_0(x,\tau)=\,{\rm sech}\,(x+\tau)$, and
the corresponding kink, $\phi_K(x,\tau)=2\arctan \tanh \frac{x+
\tau}{2}$, ends in the $\tau$-independent potential:
$U(\phi,\tau)=\frac{1}{2} \cos^2 \phi =\frac{1}{4}(1+\cos 2\phi)$.
Thus, all the members in the family are completely equivalent to the
original sine-Gordon model, although a translation in
$\frac{\pi}{2}$ and a rescaling by $2$ are needed. The alternative
(and equivalent) choice of
$\Xi_1(x,\tau)=-\log\frac{1}{\sinh\tau}\cosh(x+\tau)$ as
superpotential leads to a family of models identical to the family
proposed by Kumar, discussed above. In summ, translations and
dilatations in field space can be reinterpreted as translations and
dilatations in coordinate space, a situation encompassed in the
soldering of internal and external variables by topological defects.

To end this subsection, we mention that in correspondence
with (\ref{operator55}),  the family of isospectral kink fluctuation
operators $K_{1-}(\tau)$ can be traced back to the free particle by
means of the displaced Darboux operators
$A(\tau)=\frac{d}{dx}+\tanh(x+\tau)$ and  $A^\dagger(\tau)$, as
described in references \cite{Plyushchay2012,PlyNie,Khare1989}.
As a consequence, any pair $K_{1-}(\tau)$ and $K_{1-}(\tau')$ can be
related by the second-order intertwining operator $Y(\tau,\tau')=
A^\dagger(\tau)A(\tau')$,
$Y(\tau,\tau')K_{1-}(\tau')=K_{1-}(\tau)Y(\tau,\tau')$ coherently
with the discussion of a generic one-parametric family
(\ref{K1-Cadd}). The peculiarity of the present family of
isospectral kink fluctuation operators is, however, that any two of
them with $\tau\neq \tau'$ can also be intertwined by the first
order operator
$X(\tau,\tau')=\frac{d}{dx}-\tanh(x+\tau)+\tanh(x+\tau')
+\coth(\tau-\tau')$,
$X(\tau,\tau')K_{1-}(\tau')=K_{1-}(\tau)X(\tau,\tau')$. The product
of both, first and second order, independent intertwining operators
thus generates the non-trivial Lax integral $Z(\tau)=
A(\tau)\frac{d}{dx}A^\dagger(\tau)$ for the reflectionless Schr\"odinger
operator $K_{1-}(\tau)$:
$X(\tau,\tau')Y(\tau',\tau)=Z(\tau)+\coth(\tau-\tau') K_{1-}$,
$[Z(\tau),K_{1-}]=0$, see \cite{Plyushchay2012,PlyNie} for the
details and a discussion of the supersymmetric structure associated
with a pair $K_{1-}(\tau)$, $K_{1-}(\tau')$.

\subsection{The $\phi^4$ model kinship}

Finally, we shall consider a family of models implicitly
constructed by Kumar in \cite{Kumar1987,Dey1994}. This family is
based on the $\phi^4$ model whose second-order kink fluctuation
operator factorizes in terms of the function $W_1(x)=2\tanh x$.
Therefore, application of formula (\ref{GenericSuperpotential+})
offers the following $C$-parametric family of superpotentials and
their derivatives:
\[
\Xi_2(x,C)=\log\frac{{\rm cosh}x}{12 C  {\rm cosh}^3x-2(3{\rm
sinh}x+{\rm sinh}3x)} \, \, \, , \, \, \,  W_2(x,C)=2\tanh
x-\frac{3\,{\rm sech}^4 x}{3C-(2 + {\rm sech}^2 x) \tanh x} \, \, .
 \hspace{0.2cm}.
\]
The second-order kink fluctuation operators and their supersymmetric partners are
\begin{eqnarray}
K_{2-}(C) &=& -\frac{d^2}{dx^2}+ 4  -\frac{6 \,{\rm sech}^2 x
\,[4+9C^2-8\, {\rm sech}^2 x+ {\rm sech}^4 x+6\,C\,(-2+{\rm sech}^2
x)\tanh x]}{[-3C+(2+{\rm sech}^2 x)\tanh x]^2}  \hspace{0.4cm},\nonumber \\
K_{2+} &=& - \frac{d^2}{dx^2}+4-2\, {\rm sech}^2 x \hspace{0.4cm}.
\label{K2+-add}
\end{eqnarray}
From this we read $v_2^2=4$ and
\[
V_{2-}(x,C)= -\frac{6 \,{\rm sech}^2 x \,[4+9C^2-8\, {\rm sech}^2 x+ {\rm sech}^4
x+6\,C\,(-2+{\rm sech}^2 x)\tanh x]}{[-3C+(2+{\rm sech}^2 x)\tanh x]^2} \, \, \, ,
\]
a family of very complicated potential wells indeed. In Figure 6 the potential well of
$V_{2-}(x,C)$ has been plotted for several values of $C$. The condition $|C|>\frac{2}{3}$ must be imposed to avoid
singularities in the expression of $V_{2-}(x)$. In this range, we find that $\left< V_{2-}(x)\right>=-12$. The $K_{2+}$ operator, however, is a transparent P\"oschl-Teller Schr\"odinger operator that is identical ($C$-independent) for all the family members. The only bound state eigenvalue is $\omega_1^2=3$, whereas the phase shifts in the eigenfunctions of the continuous spectrum are: $\delta^{(2+)}(k)=2\arctan \frac{1}{k}$. The $K_{2-}(C)$ operators in turn exhibit two bound states: the zero mode, ($\omega_0^2=0$) and the SUSY paired state to the $K_{2+}$ bound state with energy (the same for all the family members) $\omega_1^2=3$. From formula (\ref{superpotential}) we identify the zero modes
and, implicitly, the kink solutions:
\begin{equation}
\psi_0(x,C)= \frac{\cosh x}{12 C\cosh^3 x-6 \sinh x-2\sinh 3x} \, \, \, , \, \, \, \phi_K(x,C)=\int \, dx \, \frac{\cosh x}{12 C\cosh^3 x-6 \sinh x-2\sinh 3x} \label{4Ckink} \, \, \, .
\end{equation}
Although (\ref{4Ckink}) tells us that the kink solutions exist for any value of $C$,  $|C|>\frac{2}{3}$, the integral in (\ref{4Ckink}) cannot in general be expressed in terms of known analytical functions and/or its computation involves expressions too complicated to be of practical use. In Figure 6, however, we have depicted the kink solutions for several values of the parameter $C$ obtained by means of numerical integration. The value $C=6$ is exceptional in the sense that the integral (\ref{4Ckink}) is feasible in terms of elementary functions and in this case we find the very explicit analytical expression for the kink solution:
\[
\phi_K(x,6) = \frac{1}{960} \left[ 6 \sqrt{15} \arctan \frac{-3+2\tanh x}{\sqrt{15}} + 10 \log[3\cosh x+\sinh x] - 5 \log
[5+7\cosh 2x - 3\sinh 2x] \right] \hspace{0.4cm}.
\]

\begin{figure}[ht]
\centerline{\includegraphics[height=3cm]{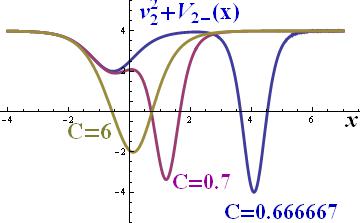} \hspace{1cm}
\includegraphics[height=3cm]{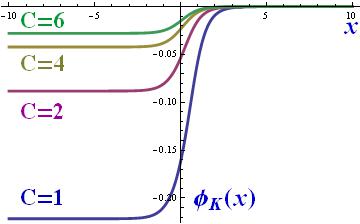} }

\caption{\small Potential wells arising in the second-order kink fluctuation operator $K_{2-}$ (left) and kink
solutions (right) for several values of $C$.}
\end{figure}

The associated potential energy densities characterizing these field theory models cannot be written explicitly because we do not know the analytical form of the kinks. Therefore, we are blind about how to invert the
kink solution. We do have, however, all the ingredients to compute the one-loop kink mass shift in these models. Formula (\ref{gdhn2}) reveals that
\[
\frac{\Delta E(\phi_K)}{\hbar\gamma_d^2}= \frac{1}{2}\sqrt{3} + \frac{1}{4\pi} (-12) + \frac{1}{2\pi} \int_0^\infty dk
\left[  -\frac{2\sqrt{k^2 + 4}}{k^2+1} - \frac{\frac{1}{2}  (-12)+ 2\cdot 2}{\sqrt{k^2 + 4}}\right] =
\frac{1}{2\sqrt{3}} - \frac{3}{\pi}\hspace{0.4cm}.
\]
Thus, all the kinks in this family undergo the same mass shift as the kink of the $\phi^4$ model.

To finish this sub-Section it is useful to write the above-discussed  family of potentials in a way that trades the $C$ parameter by spatial displacements of potential wells and kinks. The redefinition $C=-\frac{1}{3}\left({\rm tanh}\tau+{\rm coth}\tau\right)$ is similar but slightly more complicated than the analogous transformation performed
in the sub-Section 4.1. In the new form the superpotential and the function $W_2$ become
\begin{eqnarray}
\Xi_2(x,\tau)&=&\log\left[\frac{{\rm cosh}x}{3{\rm cosh}(x+2\tau)+{\rm cosh}(3x+2\tau)}\right]\nonumber
\\
W_2(x,\tau)&=& 2{\rm tanh}x +\frac{3{\rm sech}^4 x}{{\rm
coth}\tau+(2+{\rm sech}^2x){\rm tanh}x+{\rm tanh}\tau}\,
\label{W2-add}
\end{eqnarray}
The zero modes look simpler,
\[
\psi_0(x,\tau)= \frac{{\rm cosh}x}{3{\rm cosh}(x+2\tau)+{\rm cosh}(3 x+2 \tau)} \, \, \, , \, \, \, \quad \int_{-\infty}^\infty \, dx \, \psi_0^2(x,\tau)=\frac{1}{12} \, \, ,
\]
but the potential wells in the family of Hamiltonians
\begin{equation}\label{K2-tau-add}
K_{2-}(\tau)=-\frac{d^2}{dx^2}+4-\frac{12 (3+4{\rm cosh}(2x)+{\rm
cosh}[4(x+\tau)])}{\left(3{\rm cosh}(x+2 \tau)+{\rm
cosh}(3x+2\tau)\right)^2} \label{disham}
\end{equation}
show two minima with a distance between them modulated by the
translational parameter $\tau$.  In fact, this Hamiltonian is the
$\kappa_1=1$, $\kappa_2=2$, $x_1=0$, $x_2=-\tau$ member of the
family of two-gap Schr$\ddot{\rm o}$dinger operators depending on
two scaling and two translation parameters ,to be discussed in the
next sub-Section.

In a generic case, in accordance with (\ref{YCC-add}), any
pair $K_{2-}(\tau)$ and $K_{2-}(\tau')$ of the Schr\"odinger type
operators of the form (\ref{K2-tau-add}) is intertwined by the
second order operator $Y(\tau,\tau')=A_2^\dagger(\tau)A_2(\tau')$,
$Y(\tau,\tau')K_{2-}(\tau')=K_{2-}(\tau)Y(\tau,\tau')$, where
$A_2(\tau)=\frac{d}{dx}+W_2(x,\tau)$ with $W_2(x,\tau)$ given by
equation (\ref{W2-add}). Bearing in mind that the ``virtual" system
(\ref{K2+-add}) is  such that the intertwining relation can be related
to the free particle Schr\"odinger operator, the pair of
$K_{2-}(\tau)$ and $K_{2-}(\tau')$ can also be intertwined by the
fifth order operator
$X(\tau,\tau')=A_2^\dagger(\tau)A_1^\dagger\frac{d}{dx}A_1A_2(\tau')$,
$X(\tau,\tau')K_{2-}(\tau')=K_{2-}(\tau)X(\tau,\tau')$, where
$A_1=\frac{d}{dx}+\tanh x$. Putting $\tau'=\tau$ in the last intertwining
relation  we see that $X(\tau,\tau)$ is nothing else than
the nontrivial Lax integral for $K_{2-}(\tau)$,
$[X(\tau,\tau),K_{2-}(\tau)]=0$. We refer to \cite{Plyushchay2012},
where the peculiar supersymmetric structure associated with the
isospectral pair of reflectionless Schr\"odinger operators
$K_{2-}(\tau)$ and $K_{2-}(\tau')$ is discussed  in detail. In particular
it is shown there that the fifth-order operator $X(\tau,\tau')$
can be reduced to an independent intertwining operator of differential order $3$.

\subsection{From two-gap potentials to scalar field models}

Within the context of supersymmetry, in reference
\cite{Plyushchay2012}, an analysis has been made of the most general
Hamiltonian, depending on two scaling and two translation parameters,
such that the reflection scattering amplitude is zero and there are
two bound states in its spectrum:
\begin{equation}\label{K2-add}
K_{2-} =A_2^\dagger A_2 =-\frac{d^2}{dx^2}+
\kappa_2^2-2(\kappa_2^2-\kappa_1^2) \frac{\kappa_2^2 \, {\rm
csch}^2[\kappa_2(x-x_2)]+\kappa_1^2 \,{\rm sech}^2
[\kappa_1(x-x_1)]}{(\kappa_2 \coth [\kappa_2(x-x_2)]-\kappa_1
\tanh[\kappa_2(x-x_1)])^2} \quad ,
\end{equation}
where with no loss of generality we choose $\kappa_2^2
>\kappa_1^2$\, {\footnote{To cope with the opposite inequality,
$\kappa_1^2 >\kappa_2^2$, one merely exchanges the r$\hat{\rm o}$les
of $\kappa_1^2$ and $\kappa_2^2$.}}. Henceforth, we have
$v_2^2=\kappa_2^2$, $\langle V_{2-} \rangle=-4( \kappa_1 +
\kappa_2)$, and the four-parametric family of potential wells is:
\begin{equation}
V_{2-}(x,\kappa_1,\kappa_2,x_1,x_2)=  -2(\kappa_2^2-\kappa_1^2)
\frac{\kappa_2^2 \, {\rm csch}^2[\kappa_2(x-x_2)]+\kappa_1^2 \,{\rm
sech}^2 [\kappa_1(x-x_1)]}{(\kappa_2 \coth
[\kappa_2(x-x_2)]-\kappa_1 \tanh[\kappa_2(x-x_1)])^2} \label{2potw}
\, \, . \, \,
\end{equation}
This second-order operator is factorized by  means of the
first-order operator $A_2=\frac{d}{dx}+W_2(x)$ and its adjoint, where
the superpotential and its derivative are respectively
\begin{eqnarray}
 \Xi_2(x;\kappa_1,\kappa_2,x_1,x_2)&=&
 \log\left[\frac{2 \, {\rm csch}[\kappa_2(x-x_2)]}{\kappa_2
 \coth [\kappa_2(x-x_2)]-\kappa_1 \tanh
 [\kappa_1(x-x_1)]}\right]\,\,,\nonumber
 \\ W_2(x;\kappa_1,\kappa_2,x_1,x_2)&=&\frac{\kappa_2^2-
 \kappa_1(\kappa_1 \,{\rm sech}^2[\kappa_1(x-x_1)]+\kappa_2
 \coth [\kappa_2 (x-x_2)]\tanh [\kappa_1(x-x_1)])}{\kappa_2
 \coth [\kappa_2(x-x_2)] - \kappa_1 \tanh[\kappa_1(x-x_1)]}\,\,.
 \label{W2k1k2x1x2-add}
\end{eqnarray}
The ground state, read from the exponential of the superpotential, is a zero mode:
\begin{equation}
\psi_0(x;\kappa_1,\kappa_2,x_1,x_2) = \frac{2 \, {\rm csch}[\kappa_2(x-x_2)]}{\kappa_2 \coth [\kappa_2(x-x_2)]-\kappa_1 \tanh [\kappa_1(x-x_1)]} \quad , \quad \omega^2_0=0 \, \, .
\label{zeromode55}
\end{equation}
The supersymmetric partner operator is surprisingly simple and is related to the family of one-gap
operators:
\begin{equation}
K_{2+}=A_2A_2^\dagger=-\frac{d^2}{dx^2}+\kappa^2_2-2\kappa_1^2 {\rm sech}^2[\kappa_1(x-x_1)] \label{2gapsusy}\, \, .
\end{equation}
It is in fact the first member of the hierarchy of reflectionless
P\"oschl-Teller potentials with threshold of the continuous spectrum
at $\kappa_2^2$. This operator presents only a bound state
$\psi_{(\kappa_2^2-\kappa_1^2)}^+(x)={\rm sech}[\kappa_1(x-x_1)]$,
with energy $\kappa_2^2-\kappa_1^2$. One easily shows that the first
excited state of $K_{2-}$ is paired to this state by supersymmetry:
\[
\psi_1(x;\kappa_1,\kappa_2,x_1,x_2) = A_2^\dagger {\rm sech}[\kappa_1(x-x_1)] \quad , \quad \omega_1^2=\kappa_2^2-\kappa_1^2 \quad .
\]
The total phase shift is also very well known, $\delta^{(2+)}(k)=2{\rm arctan}\frac{\kappa_1}{k}$ and hence $\delta^{(2-)}(k)$ is easily obtained using supersymmetry. We remark that, unlike the case of one-gap potentials, here there is no a single superpartner operator $K_{2+}(\kappa_1, \kappa_2, x_1)$ for all the family operators $K_{2-}(\kappa_1, \kappa_2, x_1, x_2)$  (although in practice, the dependence in $\kappa_2$ of $K_{2+}$ merely shifts the energy levels globally ). Instead, what we find is a reduction from a four-parametric family of two-gap operators to a two-parametric family of one-gap operators.

Before discussing the scalar field models related to the family of $K_{2-}(\kappa_1, \kappa_2, x_1, x_2)$ second-order operators it is convenient to analyze some aspects of the family of associated potential wells $V_{2-}(x, \kappa_1, \kappa_2, x_1, x_2)$. In order to elucidate the asymptotic behavior at large values of $x_1$ and $x_2$ we write the potentials as the sum of two terms:
\[
V_{2-}(x;\kappa_1,\kappa_2,x_1,x_2)= V_{2-}^{(1)} (x;\kappa_1,\kappa_2,x_1,x_2) + V_{2-}^{(2)} (x;\kappa_1,\kappa_2,x_1,x_2)
\]
where
\begin{eqnarray*}
V_{2-}^{(1)} (x;\kappa_1,\kappa_2,x_1,x_2) &=& -\frac{2 \kappa_1^2}{(\cosh \alpha \cosh [\kappa_1 (x-x_1)] \coth [\kappa_2 (x-x_2)]-\sinh \alpha \sinh [\kappa_1 (x-x_1)])^2} \\ V_{2-}^{(2)} (x;\kappa_1,\kappa_2,x_1,x_2)&=& -\frac{2 \kappa_2^2}{(\cosh \alpha \cosh [\kappa_2 (x-x_2)]-\sinh \alpha \tanh [\kappa_1 (x-x_1)] \sinh [\kappa_2 (x-x_2))])^2} \, \, \, .
\end{eqnarray*}
 The new parameter $\alpha$ is defined in term of the old ones such that: $\tanh \alpha=\frac{\kappa_1}{\kappa_2}$. When $\kappa_2(x-x_2)\gg 0$ we have that
\[
V_{2-}^{(1)} (x;\kappa_1,\kappa_2,x_1,x_2) = - 2 \kappa_1^2 {\rm sech}^2 \Big[\kappa_1 (x-x_1)-{\rm sign} (x-x_2) \alpha\Big] \, \, \, .
\]
Analogously, for $\kappa_1(x-x_1)\gg 0$ the behavior is:
\[
V_{2-}^{(2)} (x;\kappa_1,\kappa_2,x_1,x_2)= - 2 \kappa_2^2 {\rm sech}^2 \Big[\kappa_2 (x-x_2)-{\rm sign}(x-x_1)\alpha\Big] \, \, \, .
\]
In summ, in this range of parameters the potential wells of the family resemble two one-bound state type P$\ddot{\rm o}$schl-Teller potential wells, as if they were describing two separated kinks appearing in two sine-Gordon models with different masses: $m_1^2=\kappa_1^2$ and $m_2^2=\kappa_2^2$.

The scalar field theory family is reconstructed from the kink profile identified through spatial integration of the zero modes above. It is clear that, like in the $N=1$ models of previous sections, there is an insufficient arsenal of analytical functions to achieve this task without relying on numerical integrations. Nevertheless, there are special points in the parameter space $(\kappa_1,\kappa_2, x_1, x_2)$ where analytical computation of these integrals is possible, leading to highly distinguished scalar field models.
\begin{enumerate}
\item If $\kappa_1= \frac{1}{2} \kappa_2= 1$ and $x_1=x_2=0$,
the kink profile is simply: $\phi_K(x)=\tanh x$ and we
encounter the $\phi^4$-model: $U(\phi)=\frac{1}{2}(\phi^2-1)^2$.

\item If $\kappa_1= \frac{1}{2} \kappa_2= c$,
 where $c\in\mathbb{R}$ is a real constant, and $x_1=x_2=0$, the kink profile is a dilatation of the usual kink in a re-scaled space: $\phi_K(x)=\frac{1}{c^2}\tanh (cx)$. Now, the scalar field model responds to the potential energy density: $U(\phi)=\frac{1}{2c^2}(c^4\phi^2-1)^2$, a re-scaled $\phi^4$ model.

\item If $\kappa_1=\frac{1}{2} \kappa_2=1$ and $x_1=0$, $x_2={\rm arctanh}\, \frac{-1}{9+4\sqrt{5}} $, the kink profile is
\[
\phi_K(x)= \sqrt{3} \arctan \frac{\sqrt{5}(1+2e^{2x})}{\sqrt{3}}+ \frac{\sqrt{5}}{6} \log \frac{2+5 e^{2x}+5 e^{4x}}{(2+e^{2x})^2} \quad .
\]
\item If $\kappa_1=1$, $\kappa_2=4$ and $x_1=x_2=0$, the kink profile is
\[
\phi_K(x)=\frac{1}{50}\Big[ 5 + 6\sqrt{5} \arctan \frac{-2+3e^{2x}}{\sqrt{5}} - 5 \tanh x \Big]
\]
This function and the previous one are too complicated to invert.
\item At the $x_1\rightarrow \pm \infty$ limit the kink profile becomes:
\[
\phi_K(x)= \frac{4}{\kappa_2 \sqrt{\kappa_2^2-\kappa_1^2}} \arctan \frac{\sqrt{\kappa_2\pm\kappa_1} e^{\kappa_2(x-x_2)}}{\sqrt{\kappa_2\mp\kappa_1}}\quad .
\]
 The corresponding field model obeys to a potential energy density of the form: \newline $U(\phi)=\frac{2}{\kappa_2^2-\kappa_1^2} \sin^2 \frac{\kappa_2 \sqrt{\kappa_2^2-\kappa_1^2}}{2}\, \phi$, a sine-Gordon type of scalar field model.
\end{enumerate}

In order to compute the one-loop mass shifts by using the SUSY generalized DHN formula we now pass to the next
level of the SUSY hierarchy. To start at a new level we define the operator:
\[
K_{1-}=K_{2+} +\kappa_1^2-\kappa_2^2= -\frac{d^2}{dx^2} +\kappa_1^2- 2\kappa_1^2 \, {\rm sech}^2 \,[\kappa_1(x-x_1)]= A_1^\dagger A_1 \, \, \, .
\]
$K_{1-}$ is again the first transparent  P\"oschl-Teller Hamiltonian
with the origin of energies displaced in $\kappa_1^2-\kappa_2^2$
with respect to $K_{2+}$. The first-order operators allowing the
factorization $K_{1-}=A_1^\dagger A_1$ are:
\begin{equation}\label{W1k1x1-add}
A_1^\dagger=-\frac{d}{dx}+W_1(x)  \, \, \, , \, \, \,
A_1=\frac{d}{dx}+W_1(x) \hspace{0.5cm} \mbox{where} \hspace{0.5cm}
W_1(x) = \kappa_1 \tanh [\kappa_1 (x-x_1)]
\end{equation}
Clearly $v_1^2=\kappa_1^2$ and there exists a bound state associated with the eigenvalue $\omega_1^2=\kappa_2^2-\kappa_1^2$ of $K_{2+}$, which is a zero mode of $K_{1-}$. The SUSY partner operator in this level is
\[
K_{1+} = A_1 A_1^\dagger = -\frac{d^2}{dx^2} +\kappa_1^2
\]
such that we end in the free particle operator with zero phase shift: $\delta^{(1+)}=0$. From (\ref{gdhn3}), the kink mass shift for any kink, analytically known or not, in the family can be computed to find:
\begin{eqnarray*}
\frac{\Delta E(\phi_K)}{\hbar\gamma_d^2}& =& \frac{1}{2} \sqrt{\kappa_2^2-\kappa_1^2} -  \frac{1}{\pi} (\kappa_1+\kappa_2) +\frac{1}{2\pi} \int_0^\infty dk \left[
- \frac{-2(\kappa_1+\kappa_2)+ 2\kappa_2}{\sqrt{k^2 + \kappa_2^2}}
- \frac{2\kappa_1\sqrt{k^2 + \kappa_2^2}}{k^2 + \kappa_1^2} \right] = \\ &=&  -  \frac{1}{\pi} (\kappa_1+\kappa_2) + \frac{1}{\pi} \sqrt{\kappa_2^2-\kappa_1^2} \arcsin\frac{\kappa_1}{\kappa_2} \quad .
\end{eqnarray*}
As a test, we stress that in the  $\kappa_1=1$, $\kappa_2=2$, $x_1=0$, $x_2=0$, case, which is the $\phi^4$-model, the one-loop kink mass shift $\frac{\Delta E(\phi_K)}{\hbar\gamma_d^2}=\frac{1}{2\sqrt{3}}-\frac{3}{\pi}$ is reproduced. We finally mention that this result agrees perfectly with the answer obtained using the Cahill-Comtet-Glauber formula:
\[
\frac{\Delta E^{{\rm CCG}}(\phi_K)}{\hbar\gamma_d^2}= -\frac{\kappa_2}{\pi}\sum_{i=1}^2\,(\sin \theta_i -\theta_i {\rm arccos}\theta_i) \hspace{0.5cm} , \hspace{0.5cm} \theta_1={\rm arccos}\frac{0}{\kappa_2}=\frac{\pi}{2} \hspace{0.5cm} , \hspace{0.5cm} \theta_2={\rm arccos}\frac{\sqrt{\kappa_2^2-\kappa_1^2}}{\kappa_2} \quad ,
\]
where the CCG angles are given  in terms of the bound state
eigenvalues and the continuous spectrum threshold.

To conclude this sub-Section, we note that any
isospectral pair $K_{2-}(\kappa_1,\kappa_2,x_1,x_2)$ and
$K_{2-}(\kappa_1,\kappa_2,x'_1,x'_2)$ of  reflectionless
Schr\"odinger operators of the form (\ref{K2-add}) can be related in
a generic case by two independent intertwining operators of differential
orders $3$ and $2$. We do not display their explicit
form here but simply refer to \cite{Plyushchay2012}, where the
corresponding associated supersymmetric structure can also be found.
In a special case, when the relative shift parameters $x_1-x'_1$ and
$x_2-x'_2$ are related by the equality
$\kappa_1\coth(x_1-x'_1)=\kappa_2\coth(x_2-x'_2)$, the independent
irreducible intertwining operators are, instead, the operators of
orders $4$ and $1$,
\begin{eqnarray}
Y_4(x_1,x_2,x'_1,x'_2)&=&A_2\dagger(x_1,x_2)A_1^\dagger(x_1)A_1(x'_1)A_2(x'_1,x'_2)\,\,,
\label{Y4-add}\\
X_1(x_1,x_2,x'_1,x'_2)&=&\frac{d}{dx}+W-W'-\kappa_1\coth(x_1-x'_1)
\,\,,\label{X1-add}
\end{eqnarray}
where $W=W_2(x;\kappa_1,\kappa_2,x_1,x_2)-W_1(x;\kappa_1,x_1)$ with
$W_2$ and $W_1$ defined, respectively,  in (\ref{W2k1k2x1x2-add})
and (\ref{W1k1x1-add}), while $W'$ is defined in the same way as $W$
but with $x_1$, $x_2$ replaced by $x'_1$ and $x'_2$. In particular, this means that
the second order small kink fluctuation
operator of the $\phi^4$ model with
$U(\phi)=\frac{1}{2}(\phi^2-1)^2$, to which the choice $\kappa_1=
\frac{1}{2} \kappa_2= 1$, $x_1=x_2=0$ does correspond, can be
related by exotic supersymmetry involving  the supercharges of  orders $4$ and $1$,
composed from (\ref{Y4-add}) and (\ref{X1-add}),
with the kink fluctuation operator of the form (\ref{K2-add}) with
$\kappa_1= \frac{1}{2} \kappa_2= 1$ and shift parameter $x_1$ given
in terms of the arbitrary shift parameter $x_2$ by a relation
$x_1={\rm arccotanh}(2\coth x_2)$.

\section{Conclusions and further comments}

The main conclusion to be drawn here is the fact that the intrinsic supersymmetry attached to the stability
operator of BPS kinks is helpful in the task of evaluating the one-loop mass shifts induced by kink fluctuations. Throughout this work we have
applied this idea  in increasing order of difficulty: first, we addressed very well known models from this new strategic point of view to test the method and gain confidence in its effectiveness. Second, we explored ignotum models where the new procedure showed the answer, impossible to reach by means of traditional approaches. Finally, within this new framework we have analyzed families of models with a very rich supersymmetric structure. Again, the results were extremely satisfactory.

In hindsight, these developments lend us to speculate that the extension of the method to scalar field models with more than one scalar field will pay even better dividends. Even though the heat kernel expansion allowed the computation
of BPS kink mass shifts in some interesting models, \cite{Alonso2004, Alonso2011}, there are other more interesting two-component BPS Wess-Zumino kinks, \cite{Fendley1990,Abraham1991,Alonso2000} where this computation has not yet been achieved within the purely bosonic framework. Starting with the models where some results are available we plan
to address the computation of Wess-Zumino kink mass shifts using the SUSY quantum mechanical methods developed in this paper.

Finally, we look forward to treating BPS topological defects in higher dimensions similarly, e.g. the BPS self-dual vortices in the $N=2$-dimensional Abelian Higgs model at critical coupling between Type I and II superconductivity. To accomplish this task, one must deal with supersymmetric quantum mechanical systems in $N$ dimensions. The spectral problems are thus much more involved but a fairly good quantity of information is available, see e.g. \cite{Ioffe2012} to see a recent review.

\section*{ACKNOWLEDGEMENTS}

This work has been partially supported by DICYT (USACH) and FONDECYT Grant 1095027 (Chile), and by
 the Spanish Ministerio de Educacion y Ciencia (DGICYT) under Grant FIS2009-10546.
 MP and JMG are grateful, respectively,
 to the Universidad de Salamanca  and the Universidad de Santiago de Chile
for hospitality. The three of us acknowledge to the Benasque Center of Science for providing a stimulating
scientific and natural environment.

\end{document}